\documentclass[
 reprint,
superscriptaddress,
 amsmath,amssymb,
 aps,
 prl
]{revtex4-2}

\usepackage{graphicx}
\usepackage{dcolumn}
\usepackage{bm}
\usepackage{xcolor}%

\begin{document}

\preprint{APS/123-QED}

\title{Squeezing enhanced sensing at an exceptional point}

\author{Changqing Wang}
\email{cqwang@fnal.gov}
\affiliation{Superconducting Quantum Materials and Systems Center, Fermi National Accelerator Laboratory (FNAL), Batavia, IL 60510, USA}

\author{Deyuan Hu}
\affiliation{Department of Electrical Engineering and Computer Science, The University of Michigan, Ann Arbor, Michigan 48109, USA}

\author{Silvia Zorzetti}
\affiliation{Superconducting Quantum Materials and Systems Center, Fermi National Accelerator Laboratory (FNAL), Batavia, IL 60510, USA}

\author{Anna Grassellino}
\affiliation{Superconducting Quantum Materials and Systems Center, Fermi National Accelerator Laboratory (FNAL), Batavia, IL 60510, USA}

\author{Alexander Romanenko}
\affiliation{Superconducting Quantum Materials and Systems Center, Fermi National Accelerator Laboratory (FNAL), Batavia, IL 60510, USA}

\author{Zheshen Zhang}
\email{zszh@umich.edu}
\affiliation{Department of Electrical Engineering and Computer Science, The University of Michigan, Ann Arbor, Michigan 48109, USA}

\date{\today}

\begin{abstract}
Pushing the boundaries of measurement precision is central for sensing and metrology, pursued by nonclassical resources such as squeezing, and non-Hermitian degeneracies with distinct spectral response. Their convergence, however, remains challenging. We find extraordinary enhancement of sensitivity by unifying both effects in a general framework for quantum sensing in open systems. At the parametric oscillation threshold and an exceptional point, the sensing precision exhibits a unique quartic scaling with the perturbation strength. The result generalizes to multimode squeezed-state sensors with higher-order exceptional points catered to various quantum sensing platforms.
\end{abstract}

\maketitle


\textit{Introduction.---}Quantum sensing---leveraging quantum phenomena such as entanglement and coherence to probe physical quantities---has rapidly emerged as a prominent and fast-growing field \cite{degen2017quantum,pirandola2018advances}. A long-lasting objective for quantum sensing is to boost the achievable precision in the presence of classical and quantum noise via improving measurement sensitivity and signal-to-noise ratio (SNR). Squeezing, with reduced uncertainty in one quadrature of the electromagnetic field at the expense of increasing the uncertainty in the orthogonal quadrature, constitutes a powerful resource to surpass the standard quantum limit (SQL)~\cite{loudon1987squeezed,lvovsky2015squeezed,lawrie2019quantum}. Recent years have witnessed remarkable advances in leveraging squeezing to enhance weak-signal detection across diverse quantum sensing platforms, such as LIGO \cite{tse2019quantum} and Virgo \cite{acernese2019increasing} gravitational wave detectors, nonlinear integrated photonic circuits \cite{dutt2015chip}, lithium-niobate optical resonators \cite{furst2011quantum}, mechanical oscillators \cite{wollman2015quantum}, optomechanical systems \cite{xia2023entanglement}, and superconducting cavities \cite{backes2021quantum}. In particular, operating squeezing-enhanced sensors near the parametric oscillation (PO) threshold, a phase transition point, leads to divergent behavior of susceptibility through critical quantum sensing (CQS) protocols \cite{frerot2018quantum,di2023critical,alushi2024optimality,gorecki2025interplay, alushi2025collective,beaulieu2025criticality}.

As a parallel pursuit toward ultraprecise measurements, non-Hermitian physics \cite{el2018non,feng2017non,ozdemir2019parity,wang2023non} presents new opportunities by exploiting non-Hermitian degeneracies known as exceptional points (EPs) \cite{miri2019exceptional,naghiloo2019quantum}. EPs are singularities in the parameter space, where eigenvalues and their associated eigenstates of a non-Hermitian Hamiltonian or other non-Hermitian operators become degenerate \cite{wang2021coherent,naghiloo2019quantum, chen2021quantum, kononchuk2022exceptional,wang2019non,minganti2019quantum}. Near an $n$th-order EP, a weak perturbation $\theta$ induces eigenspectral splitting scaling as $\theta^{n}$ \cite{wiersig2014enhancing,wiersig2020review}, yielding amplified responses as $\theta\rightarrow0$.
Hitherto, EP-enhanced sensing schemes have been explored in a plethora of platforms, including photonic resonators \cite{chen2017exceptional,hodaei2017enhanced,zhong2019sensing}, laser gyroscopes \cite{lai2019observation,hokmabadi2019non}, plasmonics \cite{park2020symmetry}, electronic circuits \cite{dong2019sensitive}, mechanical resonators \cite{zhang2024exceptional}, etc.

Despite these encouraging advances in EP-enhanced sensing, recent studies have sparked debates about the ability of EPs to break the fundamental precision limit imposed by quantum noise \cite{langbein2018no,ding2023fundamental,lau2018fundamental,duggan2022limitations,loughlin2024exceptional,almanakly2026probing}. Notably, the SNR of gain-assisted EP sensors is strongly limited by the Petermann factor-enhanced laser noise arising from nonorthogonal eigenstates~\cite{wang2020petermann}. Nevertheless, a quantum-noise analysis has shown EP-enhanced SNR through a quadratic scaling between sensing precision and perturbation strength for a sensor operating at the lasing threshold with a second-order EP \cite{zhang2019quantum}. Furthermore, several studies revealed intriguing interplays between squeezing and non-Hermiticity, including EP-enhanced squeezing in anti-parity-time symmetric systems \cite{luo2022quantum} and EP-assisted squeezing control in superconducting resonators \cite{teixeira2023exceptional}. Yet, key questions remain: (1) How do EPs modify the ultimate sensitivity and precision limit for quantum sensing? (2) What is the physical mechanism behind EP enhancement? (3) How do EPs and squeezing jointly influence the sensing limit? 

To address these questions, we develop a unified quantum-noise framework to clarify the role of EPs in bosonic-mode quantum sensors that concurrently generate squeezing. We evaluate the precision bound for non-Hermitian sensors composed of single or coupled squeezed bosonic modes near the PO threshold, by deriving the quantum Fisher information (QFI). Our study reveals that operating at the PO threshold with a second-order EP yields a quartic scaling between the precision limit and perturbation strength, enabling exceptional sensitivity to weak signals. We further generalize the analysis to squeezing-enhanced sensors at higher-order EPs.  

\begin{figure}[!htb] 	\centering\includegraphics[width=0.46\textwidth]{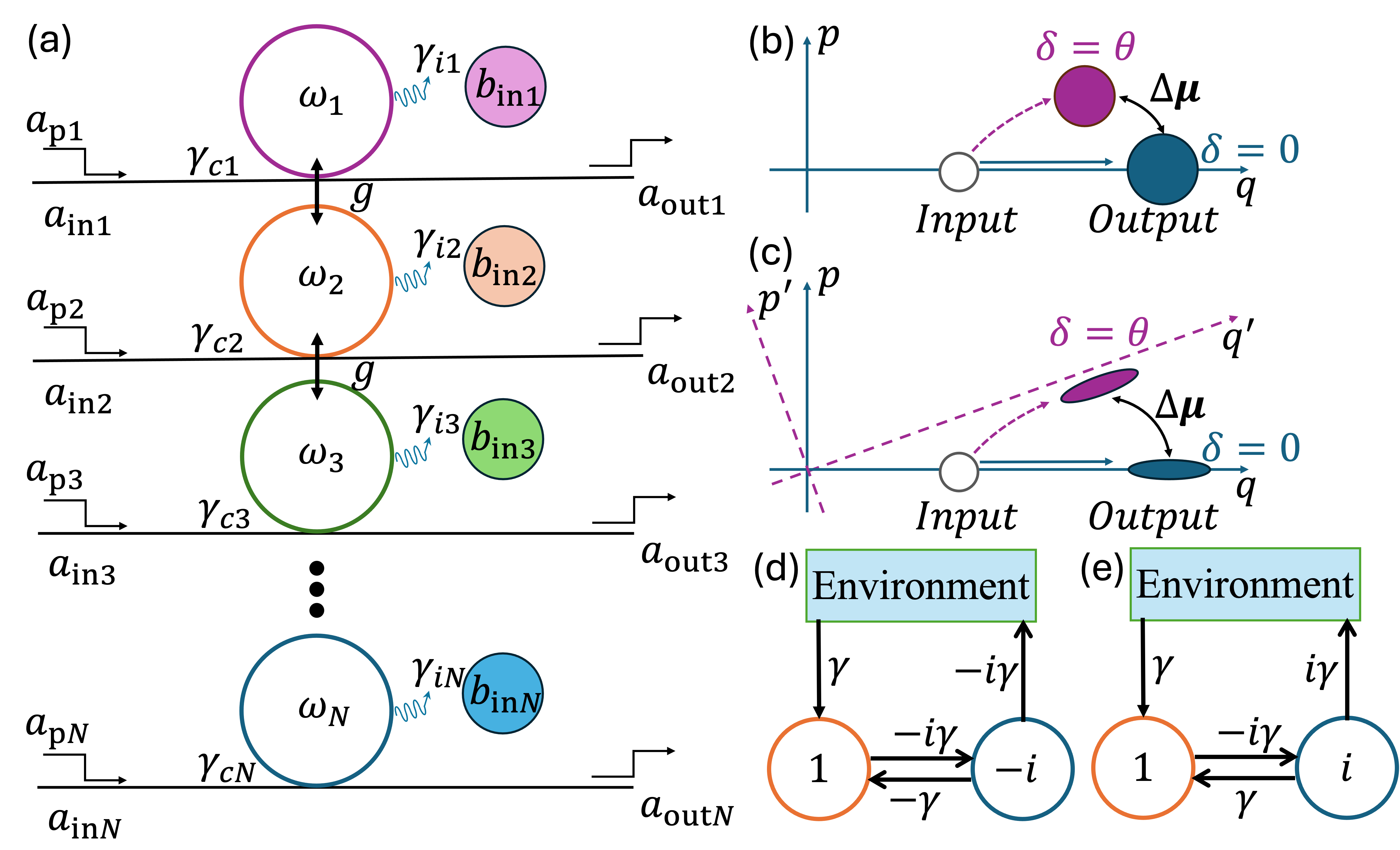}
	\caption{\normalsize Squeezing-enhanced non-Hermitian sensing. (a) Schematic diagram of a squeezing-enhanced non-Hermitian sensor consisting of $N$ coupled bosonic modes with $\chi^2$ nonlinearity. The $j$th mode with resonance frequency $\omega_{j}$ ($j=1,2,...,N$) is coupled to an input-output channel (intrinsic loss channel $b_{{\rm in}_j}$) with coupling strength $\gamma_{cj}$ ($\gamma_{ij}$). Single-mode squeezing is generated via parametric down-conversion triggered by a pump $a_{pj}$. Parameter perturbation is estimated via the scattering from the input $a_{{\rm in}_j}$ to the output $a_{outj}$. (b) Sensing near the lasing threshold. The weak perturbation $\theta$ on the resonance frequency induces a variation of the mean of output ($\Delta\bm{\mu}$) and its noise distribution represented by circles. (c) Sensing near the parametric amplification threshold. The perturbation $\theta$ rotates the eigenbasis in the quadrature space, causing a large displacement in the output ($\Delta\bm{\mu}$). (d)(e) Energy flow at an exceptional point. Two modes with amplification (orange circle) and dissipation (blue circle) each at rate $\gamma$ are coupled with strength $\gamma$. (d) For the eigenstate $(1,-i)^T$, destructive interference between the modal coupling channel and the gain/loss in each mode yields steady energy. (e) For a noneigenstate such as $(1,i)^T$, the interference becomes constructive, producing energy growth during evolution.}
	\label{Fig1}
\end{figure}

\textit{Non-Hermitian bosonic sensors with squeezing.---}We consider a canonical bosonic sensing system consisting of $N$ coupled  harmonic oscillators, each supporting a bosonic mode $a_{j}$ at the resonance frequency $\omega_{j}$ ($j=1,2,...,N$) [Fig.~\ref{Fig1}(a)]. A weak external signal imparts a uniform perturbation $\theta$, shifting the resonance frequencies as $\omega_j$ $\rightarrow$ $\omega_j+\theta$. We assume that each cavity supports degenerate parametric amplification that produces single-mode squeezing of amplitude $\epsilon_j$ through $\chi^2$ nonlinearity driven by a coherent pump field at $\omega_{pj}$. Furthermore, each mode is coupled to an intrinsic loss channel $b_{j}$ at rate $\gamma_{0j}$, and to an external input/output channel $a_{{\rm in}_j}$ (such as a waveguide) with coupling strength $\gamma_{cj}$. To incorporate phase-insensitive amplification, such as optical gain, we include coupling to an auxiliary scattering channel $c_{j}$ with strength $\kappa_j$. One can describe the system using a generic Hamiltonian under the rotating frame

\begin{align}
H = \Sigma_j& \hbar (\delta_j + \theta) a_j^\dagger a_j + \Sigma_{j,k}\hbar g (a_k^\dagger a_j + \text{H.c.}) \nonumber\\
&+ \Sigma_j\frac{i \hbar}{2} (\epsilon_j a_j^{\dagger 2} - \epsilon_j^* a_j^2),
\end{align}
where $\delta_{j}=\omega_{j}-\omega_{pj}/2$. 

Given a coherent probe, the unknown parameter $\theta$ is estimated via quadrature measurement of the output field. We define the quadratures as $q_j = (a_j + a_j^\dagger)$ and $p_j = -i(a_j - a_j^\dagger)$. In the frequency domain, the system response is characterized by the Green's function \( \bm{G}_{\theta}[\omega]\) expressed in the quadrature basis
\begin{equation}
\bm{G}_{\theta}[\omega] = -(\omega \bm{I} - \bm{M})^{-1} (\bm{I}_N \otimes \bm{\Omega}),
\end{equation}
where $\bm{M}$ is the Hamiltonian in the quadrature basis, $\bm{I}_N$ is an $N$-by-$N$ identity matrix, and $\bm{\Omega} = \begin{pmatrix} 0, 0, 1, 0; 0, 0, 0, -1; -1, 0, 0, 0; 0, 1, 0, 0 \end{pmatrix}$.

In absence of squeezing, the output is a coherent state with isotropic quadrature noise distribution in the $(q,p)$ plane. Phase-insensitive amplification amplifies the signal and noise equally, leaving the SNR of the output unchanged [Fig.~\ref{Fig1}(b)]. Nevertheless, the estimate accuracy improves since the same perturbation induces a larger excursion in the output. Amplifying the perturbation-to-output transduction is the core mechanism behind sensitivity enhancement under threshold conditions \cite{zhang2019quantum}. 

On the other hand, squeezing reshapes the noise distribution anisotropically, enabling phase-sensitive amplification to surpass the SQL. Combining squeezing and the threshold condition leads to our first essential discovery, i.e., the sensitivity is prominently enhanced when a mode operates near the PO threshold, where the dissipation is balanced by the squeezing-induced amplification. It relies on the phenomenon that approaching the PO threshold generates a giant displacement of mean ($\bm\mu$) and amplification of noise along one quadrature (for example, $q$), whereas a constant mean and uncertainty are associated with the orthogonal quadrature ($p$) [Fig.~\ref{Fig1}(c)]. The tiny perturbation causes a rotation of the squeezing orientation, i.e., the eigenbasis, from $(q,p)$ to $(q^{'},p^{'})$, leading to a new output state with the same coherent probe. The difference between the means of the two outputs ($\Delta\bm{\mu}$) is amplified by the large displacement in the antisqueezing direction. Measuring $\Delta\bm{\mu}$ via the squeezed quadratures with minimum noise yields enhanced SNR. 

\textit{Precision bound of a single-mode sensor.---}We first analyze a single-mode bosonic sensor using the quantum noise model to elucidate the sensing enhancement at the threshold, as compared to conventional sensing schemes. There exists a smallest achievable standard deviation that quantifies the lower bound of the sensing precision, known as the Cramér-Rao bound
\begin{equation}
    \Delta \theta_{\text{CRB}} \geq \frac{1}{\sqrt{N_m} \sqrt{\mathcal{I}(\theta)}},
\end{equation}
where $N_m$ is the number of measurement rounds, and $\mathcal{I}(\theta)=\mathcal{I}_{\mu}(\theta)+\mathcal{I}_{V}(\theta)$ is the QFI that characterizes the sensitivity of measurement, which generalizes the conventional scalar SNR. For sensing schemes measuring a Gaussian state with mean $\bm{\mu}_{\theta}$ and covariance $\bm{V}_{\theta}$, $\mathcal{I}_{V}(\theta)$ depends solely on $\bm{V}_{\theta}$ (see Sec.~II.C of Supplemental Material~\cite{suppl}) \cite{jiang2014quantum, zhang2019quantum}. For sufficiently large $\bm{\mu}_\theta$, QFI is dominated by $\mathcal{I}_{\mu}(\theta)$ given by 
\begin{align}
\mathcal{I}_{\mu}(\theta) &= \left( \frac{d\, \bm{\mu}_\theta}{d\theta} \right)^T 
\bm{V}_\theta^{-1} 
\left( \frac{d\, \bm{\mu}_\theta}{d\theta} \right).
\end{align}
 
\begin{figure}[!htb] 	\centering\includegraphics[width=0.46\textwidth]{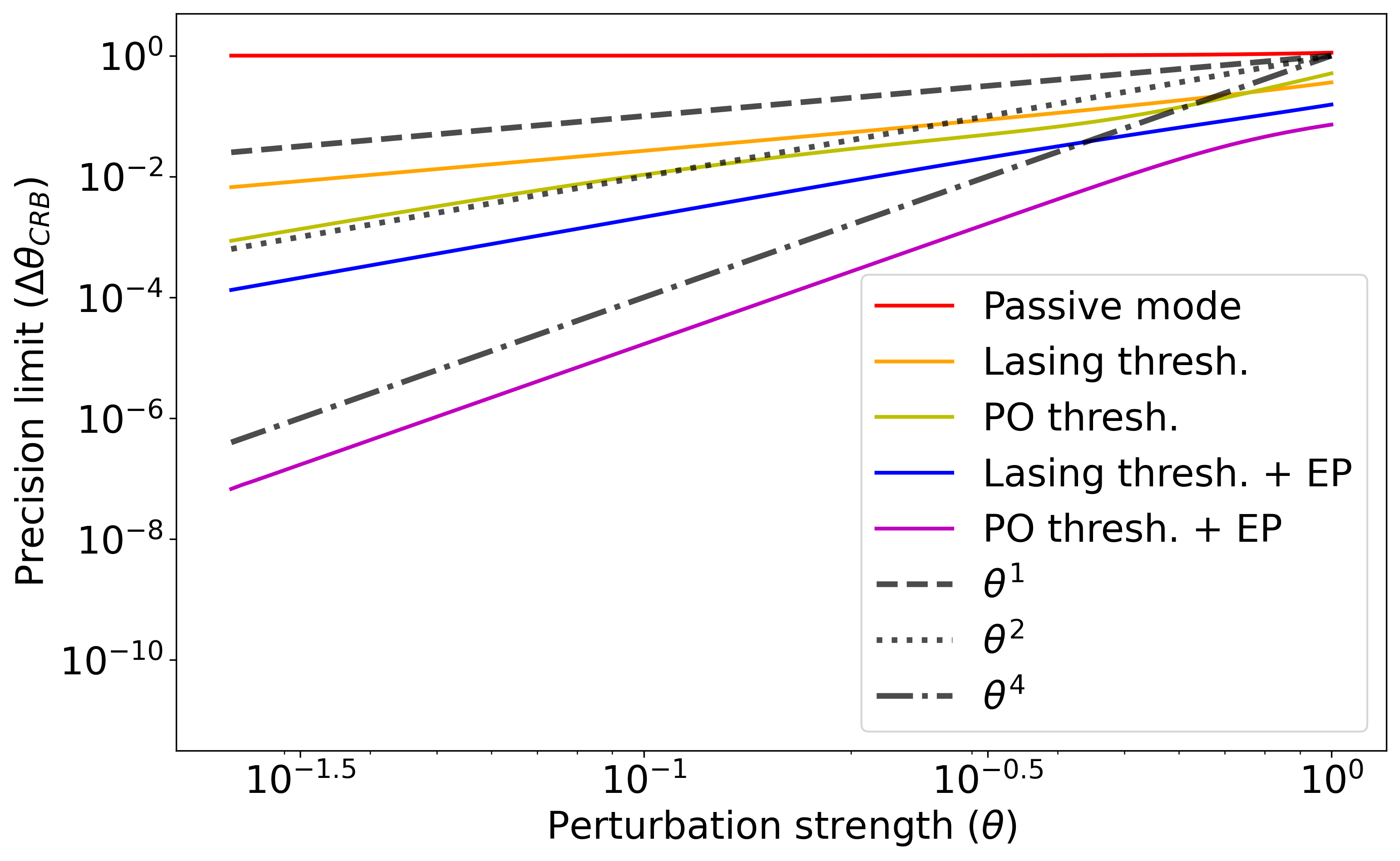}
	\caption{\normalsize Precision limit (Cramér-Rao bound $\delta\theta_{\rm CRB}$) determined by QFI, as a function of the perturbation strength ($\theta$) for different cases: a single passive mode, a single mode at the lasing threshold, a single mode at the PO threshold, two modes at an EP and the lasing threshold, and two modes at an EP and the PO threshold. The dashed, dotted, and dash-dotted lines represent the linear, quadratic, and quartic scaling, respectively. }
	\label{Fig2}
\end{figure}

For a passive single-mode sensor, the Green's function becomes $\theta$ independent for small perturbation. Consequently, the Cramér-Rao bound saturates as $\theta\rightarrow 0$ (red curve in Fig.~\ref{Fig2}). Nevertheless, one can surpass this precision limit through phase-insensitive amplification, since the diverging behavior of \( \bm{G}_{\theta}[\omega]\) boosts $d\bm{\mu}/d\theta$. At the lasing threshold, where the total loss is balanced by the gain ($\gamma_{0} + \gamma_{c} - \kappa = 0$), $\textbf{G}_{\theta}\propto\theta^{-1}$ as $\theta\rightarrow 0$. Hence QFI $\propto \theta^{-2}$ and $\delta \theta_{\text{CRB}} \propto \theta$ (orange curve in Fig.~\ref{Fig2}). 

With squeezing, the Hamiltonian in the quadrature basis (denoted as $\bm{M}$) has eigenvalues
\(\lambda_{1,2}=-i\frac{\gamma}{2} \pm \sqrt{\delta^2-|\epsilon|^2}\),  corresponding to an amplified and a deamplified eigenstate in the regime $\left|\delta\right|\leq\left|\epsilon\right|$, with the other pair $\lambda_{3,4}=-\lambda_{1,2}$. The PO threshold is reached when $\lambda_{1,3}$ become purely real, with the condition $\delta = \pm\sqrt{|\epsilon|^2 - \frac{\gamma^2}{4}}$. For $\delta=0$ and weak perturbation $\theta\ll|\epsilon|$, we have
\begin{equation}
    \lambda_{1,2} = -i\frac{\gamma}{2} \pm i|\epsilon| 
    \left( 1 - \frac{\theta^2}{2|\epsilon|^2} \right),
\end{equation}
indicating a $\theta^2$ scaling for $\lambda_{1,3}$ near the PO threshold ($|\epsilon| = \frac{\gamma}{2}$). Thus $\bm{G}_{\theta}$ contains elements scaling as $\theta^{-2}$, which ultimately yields a precision bound
$\delta\theta_{\text{CRB}}  \propto \theta^{2}$ (green curve in Fig.~\ref{Fig2}).

\textit{Precision bound of a coupled-mode sensor---}We now clarify the role of EPs and their joint effects with squeezing using a coupled-mode model. In the previously-discussed cases, if one merely increases the number of bosonic modes without intermodal coupling, the Cramér-Rao bound retains the same scaling as in the single-mode case. In contrast, a coupled-mode sensor operating at an $n$th-order EP and the lasing threshold exhibits a $\theta^n$ scaling for the Cramér-Rao bound \cite{zhang2019quantum}. Yet the physical origin of the EP enhancement has not been fully clarified. 

Using the quadrature-basis model, we identify the mechanism behind the EP-enhanced sensitivity near an amplification threshold. For phase-insensitive gain, the quadratures at $\omega$ and $-\omega$ are decoupled, enabling a simplified 4-by-4 Hamiltonian description. In particular, at a second-order EP and the lasing threshold, the effective Hamiltonian under perturbation is $\bm{M}_{\theta}=\theta \bm{I_4}-\bm{I_2}\otimes\bm{M}_{\text{EP}}$, where $\bm{M}_{\text{EP}}$ is a 2×2 Jordan block $(0,1;0,0)$ \cite{zhang2019quantum}. The non-Hermitian degenerate eigenstate $(1,-i)^T$ for $\bm{M}_{\text{EP}}$ evolves with conserved energy since in each oscillator there is a destructive interference between the incoming wave via the modal coupling channel and the gain/loss through environment [Fig.~\ref{Fig1}(d)]. Under probe, the dominant order of the system response is determined by $\bm{M}_{\theta}^{-1}=\bm{I}_2\otimes(\theta^{-1}, -\theta^{-2}; 0, \theta^{-1})$. Along the eigenstate, the response scales as $\theta^{-1}$, similar to a single-mode sensor at the lasing threshold; yet a probe misaligned with the eigenstate produces a $\theta^{-2}$ response captured by the off-diagonal term. This higher-order scaling arises because exciting a noneigenstate breaks the gain-loss balance, causing net energy absorption from environment and amplification of the intracavity field. An example is shown in Fig.~\ref{Fig1}(e), where a constructive interference occurs for input that excites $(1,i)^T$ which is orthogonal to the eigenstate. Thus, an EP further enhances the response of a sensor at the lasing threshold. Such enhancement is absent in conventional EP sensing schemes that solely rely on perturbation-induced eigenspectral splitting, or that operate away from the lasing threshold, where the added noise scales as rapidly as the signal. 

To extend EP enhancement to squeezed-mode sensors, we bring two or multiple modes with squeezing to an EP, generating a configuration where all the amplified quadratures reach a non-Hermitian degeneracy through intricate balance of their coupling, dissipation, and amplification; so do all the deamplified quadratures. To estimate the sensing precision, we first consider a two-mode scheme described by the quantum Langevin equations
\begin{align}
\frac{d a_{1,2}}{dt} = & (-i \delta_{1,2} - \frac{\gamma_{1,2}}{2}) a_{1,2} - i g a_{2,1} + \epsilon_{1,2} a_{1,2}^\dagger \\&+ \sqrt{\gamma_{c1,c2}} a_{{\rm in}_1, {\rm in}_2} \nonumber + \sqrt{\gamma_{01,02}} b_{1,2} + \sqrt{\kappa_{1,2}} c_{1,2}^\dagger,\nonumber\\
\frac{d a_{1,2}^\dagger}{dt} = & (i \delta_{1,2} - \frac{\gamma_{1,2}}{2}) a_{1,2}^\dagger + i g a_{2,1}^\dagger + \epsilon_{1,2}^* a_{1,2} \\
&+ \sqrt{\gamma_{c1,c2}} a_{{\rm in}_1, {\rm in}_2}^\dagger \nonumber + \sqrt{\gamma_{01,02}} b_{1,2}^\dagger + \sqrt{\kappa_{1,2}} c_{1,2}.\nonumber
\end{align} Single-mode squeezing in each mode couples $a_j$ and $a_{j}^\dagger$ ($j=1,2$), thereby mixing the quadratures at $\omega$ and $-\omega$, in contrast to the case with phase-insensitive amplification where the two sidebands are decoupled. Transforming to the frequency domain, $a_j \rightarrow a_j[\omega]$, and defining the quadratures as $\hat{q}_{j}[\omega] = \hat{a}_{j}[\omega] + \{\hat{a}_{j}[\omega]\}^{\dagger}, \quad \hat{p}_{j}[\omega] = -i\big(\hat{a}_{j}[\omega] - \{\hat{a}_{j}[\omega]\}^{\dagger}\big)$, one can derive an 8-by-8 non-Hermitian Hamiltonian matrix in the quadrature basis $\{ \hat{q}_1[\omega], \hat{q}_1[-\omega], \hat{p}_1[\omega], \hat{p}_1[-\omega], \hat{q}_2[\omega], \hat{q}_2[-\omega], \hat{p}_2[\omega], \hat{p}_2[-\omega] \}^T$. and the corresponding Green's function \(\bm{G}_{\theta}[\omega] = -(\omega \bm{I} - \bm{M}_{\theta})^{-1} (\bm{I}_{2} \otimes\bm{\Omega})\).
Additionally, if phase-insensitive amplification is present, additional noise is introduced from the amplification channel with the scattering described by a secondary Green's function $\bm{G}_{\theta}^{'}[\omega] = -(\omega \bm{I} -\bm{M}_{\theta})^{-1} (\bm{I}_{2} \otimes \bm{\Omega^{'}})$, where \(\bm{\Omega^{'}} = (0, 0, 0, 1; 0, 0, -1, 0; 0, 1, 0, 0; -1, 0, 0, 0)\).

We model the input field vector with a mean amplitude $\bm{\mu}_{\rm in}$ and a covariance matrix $\bm{V}_{\rm in}$, while the noise from the dissipation and amplification channels has zero means and covariance matrices $\bm{V}_{\rm in}^{'}$ and $\bm{V}_{\rm in}^{''}$, respectively. The scattering through the Gaussian channel yields an output field with the mean and covariance matrix given as
\begin{align}
    \bm{\mu}_{\rm out} &= (\bm{I}- \bm{K}_{\rm ex} \bm{G}_{\theta} \bm{K}_{\rm ex})\bm{\mu}_{\rm in},\\    
\bm{V}_{\rm out}&=(\bm{I}-\bm{K}_{\rm ex} \bm{G}_{\theta}\bm{K}_{\rm ex})\bm{V}_{\rm in}(\bm{I}-\bm{K}_{\rm ex} \bm{G}_{\theta}\bm{K}_{\rm ex})^{T}\\
&+\bm{K}_{\rm ex}(\bm{G}_{\theta}\bm{K}_{i}\bm{V}_{\rm in}^{'}\bm{K}_{i}^T\bm{G}_{\theta}^T+\bm{G}_{\theta}^{'}\bm{K}_{g}\bm{V}_{\rm in}^{''}\bm{K}_{g}^T\bm{G}_{\theta}^{'T})\bm{K}_{\rm ex}^T,\nonumber
\end{align}
where \(\bm{K}_{\mathrm{ex}}=\mathrm{diag}(\sqrt{\gamma_{c1}},\sqrt{\gamma_{c2}})\otimes\bm{I}_{4}\),\(\quad
\bm{K}_{i}= \mathrm{diag}(\sqrt{\gamma_{i1}},\sqrt{\gamma_{i2}})\otimes\bm{I}_{4}\), and \(\quad
\bm{K}_{g}= \mathrm{diag}(\sqrt{\kappa_{1}},\sqrt{\kappa_{2}})\otimes\bm{I}_{4}\).

When operating the sensor at the PO threshold, one can reach a second-order EP by tailoring the parameters to engineer pairwise degenerate eigenstates (see Sec.~IV.A of Supplemental Material~\cite{suppl} for specific conditions). There exists a $\bm{P}$ matrix that rotates $\bm{M}$ to the form \(\tilde{\bm{M}} ={\bm{P}}^{-1} {\bm{M}} \bm{P} =  \bm{M}_{\rm EP} + \mathrm{diag}\{a\theta^2, a\theta^2, -a\theta^2, -a\theta^2, b+a\theta^2, b+a\theta^2, -b-a\theta^2, -b-a\theta^2\}\),
where $a$ and $b$ are constants and $\bm{M}_{\rm EP} = \bm{I_4}\otimes\left(0, 1; 0, 0\right)$ contains four Jordan blocks. The $\theta^2$ scaling of the first four diagonal entries provides the sensitivity enhancement, as shown below.

For a general $n$th-order EP sensor at the PO threshold, $\bm{G}_{\theta}[\omega]$ can be evaluated using the Taylor expansion
\begin{align}
&\bm{G}_{\theta}[\omega] = \left( \omega \bm{I} - {\bm{P}} {\bm{\tilde{M}}} \bm{P}^{-1} \right)^{-1} (\bm{I}_{n} \otimes \bm{\Omega}) \nonumber\\
\approx & \bm{P} \left( \begin{array}{ccc} -(-a)^{-n} \theta^{-2n} \bm{M}_{\text{EP},n}^{n-1} & \bm{0} & \bm{0} \\
\bm{0} & -a^{-n} \theta^{-2n}\bm{M}_{\text{EP},n}^{n-1}  & \bm{0}\\
\bm{0} & \bm{0} & \bm{F}_n
\end{array} \right) \bm{P}^{-1}\nonumber\\
&\times(\bm{I}_{n} \otimes \bm{\Omega}),
\end{align}
where $\bm{F}_n=\text{diag}\{[(b + a\theta^2) \bm{I}_n + \bm{M}_{\text{EP},n}]^{-1}, [(-b - a\theta^2) \bm{I}_n + \bm{M}_{\text{EP},n}]^{-1}\}$. The higher-order series are terminated by $\bm{M}_{\rm EP}^n = 0$ for the nth-order EP at $\omega=0$. 
We thus obtain QFI $\sim \theta^{-4n}$ and $\delta\theta_{\rm CRB}  \sim \theta^{2n}$ for $\theta\rightarrow 0$, in contrast to the precision scaling $\delta\theta_{\rm CRB}  \sim\theta^{n}$ for EP sensors at the lasing threshold in absence of squeezing \cite{zhang2019quantum}.

\begin{figure}[!htb] 	\centering\includegraphics[width=0.46\textwidth]{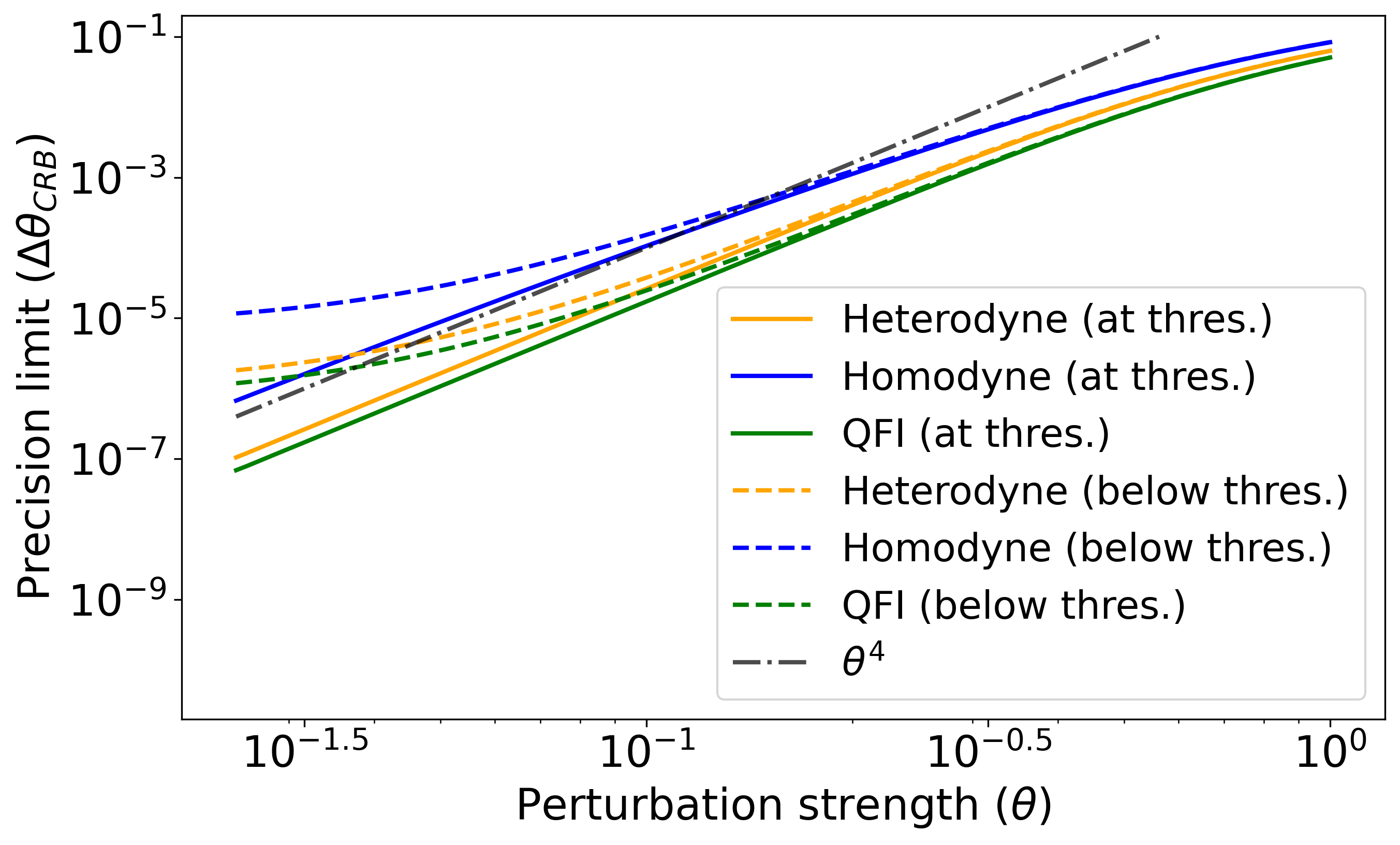}
	\caption{\normalsize Precision lower bound $\Delta\theta_{CRB}$ for an EP sensor at (solid curves) and slightly below (dashed curves) the PO threshold versus perturbation strength $\theta$, derived from QFI, as well as from the classical quantum information (CFI) achieved by heterodyne and homodyne detection. Parameters: $g=0.1$, $\epsilon_1=2i$, $\epsilon_2=-2i$, $\gamma_{c1} = \gamma_{c2} = 1$. For the PO threshold: $\gamma_{01}=2.8$, $\gamma_{02}=3.2$. For below the PO threshold: $\gamma_{01}=2.801$,$\gamma_{02}=3.201$.}\label{Fig3}
\end{figure}

This precision advantage originates from the nontrivial interplay between EPs and squeezing. If the two mechanisms acted independently, their contributions would be separable, yielding a $\theta^{n+2}$ scaling in the precision, which is analogous to sequentially passing a probe through an EP sensor and a squeezing-enhanced sensor. Instead, the $\theta^{2n}$ scaling indicates that the squeezing and EP effects are intrinsically coupled rather than additive, as confirmed by the third-order EP sensing case (see Sec.~IV.B of Supplemental Material~\cite{suppl}).

For a second-order EP sensor at the PO threshold, we have $\delta\theta_{\rm CRB} \sim \theta^{4}$ (purple curve in Fig.~\ref{Fig2}), which outperforms other conventional schemes, including EP sensors at the lasing threshold (blue curve in Fig.~\ref{Fig2}). Hence, the joint effects of squeezing and EPs provide a remarkable sensitivity advantage over conventional approaches. 

\textit{Classical Fisher information for squeezing-enhanced EP sensing.---}The QFI-bounded precision limit can be approached through homodyne or heterodyne detection in realistic measurement. The precision lower bound is evaluated by the CFI, with the dominant term $I_{\mu}(\theta)= \left( \frac{d\bm{\mu}_\theta}{d\theta} \right)^T 
\bm{\Sigma}_\theta^{-1} 
\left( \frac{d\bm{\mu}_\theta}{d\theta} \right)$ for large input, where $\bm{\Sigma}_{\theta}$ denotes the total noise covariance matrix (see Sec.~V of Supplemental Material~\cite{suppl} for explicit forms). Figure~\ref{Fig3} shows that heterodyne or homodyne detection yields a $\theta^4$ precision scaling, consistent with the QFI-limited Cramér-Rao bound. Notably, the scaling is invariant to external attenuation in realistic setups such as transmission-line dissipation or detector inefficiency. 

\textit{Effect of imperfect PO threshold.}---For EP sensors that operate slightly below the PO threshold, the scaling is cutoff for small $\theta$. Fig.~\ref{Fig3} displays the CFI and QFI with additional losses added to an EP sensor at the PO threshold. The precision limit no longer exhibits a quartic dependence on $\theta$; nevertheless, it still outperforms the conventional schemes if the imbalanced loss is much smaller than $\theta$ (see Sec.~VI of Supplemental Material~\cite{suppl}).

\textit{Discussion.---}By developing a unified framework that integrates EPs and single-mode squeezing, we find that the precision limit scales as the $2n$th power of the perturbation strength at the PO threshold for an $n$th-order EP. The results clarify how EPs enhance the fundamental detection limit of sensors combined with phase-sensitive amplification under noise constraints. 

A relevant configuration operating near the PO threshold is the CQS scheme \cite{chu2021dynamic}. In the static CQS protocol, the system is adiabatically tuned to the critical point and  measured immediately after a perturbation. In the dynamic CQS protocol, the system evolves from the vacuum state for a finite time before the measurement. Both yield QFI near a Heisenberg limit, scaling quadratically with the photon number ($\mathcal{I}\propto N^2$). In comparison, for an nth-order EP sensor at the PO threshold, QFI scales linearly with the photon number, since $\bm{G}_{\theta}\propto\theta^{-2n}$(see Sec.~IV.C.a of Supplemental Material~\cite{suppl}). Yet the precision bounds exhibit a $\theta^{2n}$ scaling, with no analog in existing CQS protocols. Thus, integrating EPs and squeezing offers a superior precision for detecting ultraweak perturbations, and relies solely on linear steady-state response to a coherent probe, without the time and procedures for state preparation or evolution.   

The proposed scheme is compatible with a broad range of experimental platforms, notably nonlinear photonic devices (see Sec.~VII of Supplemental Material~\cite{suppl} for details) and superconducting circuits. Optical squeezing is produced via parametric down-conversion \cite{wu1987squeezed} or degenerate four-wave mixing in microring resonators \cite{wu2020quantum,slusher1985observation,dutt2016tunable}, while EPs can be realized by tuning resonance frequencies and intermodal coupling \cite{wang2021coherent,wang2020electromagnetically,chen2017exceptional}. In circuit-QED systems, microwave squeezing arises from Josephson-mediated parametric amplification \cite{clerk2010introduction}, and non-Hermitian dynamics can be engineered through pulse control, flux bias, and piezoelectric positioning \cite{partanen2019exceptional}. Stabilization of EPs near the threshold can be facilitated by real-time feedback and automated parameter control.  

\begin{acknowledgments}
We thank Liang Jiang for helpful discussions. 
This work was supported by the U.S. Department of Energy, Office of Science, National Quantum Information Science Research Centers, Superconducting Quantum Materials and Systems Center (SQMS), under Contract No. 89243024CSC000002. Fermilab is operated by Fermi Forward Discovery Group, LLC under Contract No. 89243024CSC000002 with the U.S. Department of Energy, Office of Science, Office of High Energy Physics. Z.Z. also acknowledges the Office of Naval Research MURI Award No. N000142612102 and the National Science Foundation CAREER Award Grant No. 2317471.
\end{acknowledgments}

\appendix

\nocite{*}

\bibliographystyle{apsrev4-2}
\bibliography{references}

\end{document}


\title{Squeezing Enhanced Sensing at an Exceptional Point: Supplemental Material}

\author{Changqing Wang}
\email{cqwang@fnal.gov}
\affiliation{Superconducting Quantum Materials and Systems Center, Fermi National Accelerator Laboratory (FNAL), Batavia, IL 60510, USA}

\author{Deyuan Hu}
\affiliation{Department of Electrical Engineering and Computer Science, The University of Michigan, Ann Arbor, Michigan 48109, USA}

\author{Silvia Zorzetti}
\affiliation{Superconducting Quantum Materials and Systems Center, Fermi National Accelerator Laboratory (FNAL), Batavia, IL 60510, USA}

\author{Anna Grassellino}
\affiliation{Superconducting Quantum Materials and Systems Center, Fermi National Accelerator Laboratory (FNAL), Batavia, IL 60510, USA}

\author{Alexander Romanenko}
\affiliation{Superconducting Quantum Materials and Systems Center, Fermi National Accelerator Laboratory (FNAL), Batavia, IL 60510, USA}

\author{Zheshen Zhang}
\email{zszh@umich.edu}
\affiliation{Department of Electrical Engineering and Computer Science, The University of Michigan, Ann Arbor, Michigan 48109, USA}

\date{\today}

\maketitle
\setcounter{secnumdepth}{2}

\section{Introduction}

We consider a system consisting of (1) one cavity/resonator that has single-mode squeezing or (2) two or multiple coupled cavities/resonators with single-mode squeezing at the same frequency. For each scheme, we analyzed the scattering properties under quantum noise model \cite{gardiner2004quantum}, and evaluate the lower bound of sensing precision by calculating the Cramér-Rao bound determined by the quantum Fisher Information (QFI) \cite{degen2017quantum}. In particular, we investigate the exceptional point (EP)-enhanced sensors with single-mode squeezing near the parametric oscillation (PO) threshold. Furthermore we discuss the precision limit for homodyne and heterodyne detection, as well as the effect of imperfect PO threshold. Finally, we address the practical considerations relevant to experimental implementation. 

\section{Single-mode bosonic sensor}
We first study the non-Hermitian effect in a single-mode bosonic system with concurrently generated squeezing in the cavity/resonator. For generality, we also include dissipation and phase-insensitive amplification (gain) in our model.

\subsection{Quantum dynamics}
We consider a bosonic mode with resonance frequency $\omega_0$. A pump with frequency $\omega_p$ triggers the single-mode squeezing with an amplitude $\epsilon$. Under the rotating frame that makes $\epsilon$ time-invariant, the system can be described by the Hamiltonian
\begin{equation}
    H = \hbar\delta a^{\dagger} a
    +\frac{i \hbar}{2}\left( \epsilon a^{\dagger 2} -\epsilon^{*} a^2\right),
\end{equation}
where $\delta=\omega_0-\omega_p/2$. The evolution of $a$ and $a^{\dagger}$ follows the dynamic equations
\begin{align}
    \frac{da}{dt}&=(-i\delta-\frac{\gamma}{2}) a + \epsilon a^{\dagger} + \sqrt{\gamma_c} a_{\rm in}+\sqrt{\gamma_0} b_{\rm in}-\sqrt{\kappa} c_{\rm in}^{\dagger},\label{EM1}\\
    \frac{da^{\dagger}}{dt}&=(i\delta-\frac{\gamma}{2}) a^{\dagger} + \epsilon^* a + \sqrt{\gamma_c} a_{\rm in}^{\dagger}+\sqrt{\gamma_0} b_{\rm in}^{\dagger}-\sqrt{\kappa} c_{\rm in},\label{EM2}
\end{align}
where the system is coupled to an intrinsic loss channel ($b_{\rm in}$) by the coupling rate $\gamma_0$, to a coupling channel ($a_{\rm in}$) by the coupling rate $\gamma_c$, and to an amplification channel ($c_{\rm in}$) by the gain rate $\kappa$. 
$\gamma=\gamma_0+\gamma_c-\kappa$ is the net loss of the mode. One can then write a dynamic matrix in the rotating frame as an effective non-Hermitian Hamiltonian \cite{wang2019non}
\begin{equation}
    M= \begin{bmatrix}
    \delta-i\frac{\gamma}{2} & i\epsilon\\
i\epsilon^* & -\delta-i\frac{\gamma}{2}
\end{bmatrix}.
\end{equation}
We can find the eigenvalues of the non-Hermitian Hamiltonian which coincide with the poles of the scattering matrix \cite{wang2021coherent}
\begin{equation}
    \lambda_{\pm}=-i\frac{\gamma}{2} \pm \sqrt{\delta^2-|\epsilon|^2},
\end{equation}
which indicates the presence of an EP around $\delta = |\epsilon|$. The lasing threshold, however, is not achieved for the eigenmode at an EP, unless the gain and loss are fully balanced ($\gamma_0 + \gamma_c = \kappa$ and $\gamma=0$). 

In the regime where $\delta\leq\epsilon$, single-mode squeezing, which is a type of phase-sensitive amplification, leads to an amplified mode (with an eigenvalue $\lambda_{+}=-i\frac{\gamma}{2} +i \sqrt{|\epsilon|^2-\delta^2}$) and a de-amplified mode (with an eigenvalue $\lambda_{-}=-i\frac{\gamma}{2} -i \sqrt{|\epsilon|^2-\delta^2}$). For nonzero $\epsilon$, we define the PO threshold as the condition at which the eigenvalue associated with the amplified mode becomes purely real. Therefore, if $\delta=0$, the PO threshold is reached by fully balancing the net loss with squeezing-induced amplification, i.e., $|\epsilon| = \frac{\gamma}{2}$. If $\delta \neq 0$, the threshold condition becomes
\begin{equation}
    \delta = \sqrt{|\epsilon|^2 - \frac{\gamma^2}{4}}.
\end{equation}

From now on, we assume a perturbation is applied and shifts the resonance frequency, i.e., $\delta\rightarrow\delta+\theta$. If the resonance-pump detuning $\delta$ is equal to 0, and the perturbation $\theta$ is much smaller than the squeezing amplitude $|\epsilon|$, the eigenvalues are approximately
\begin{equation}
    \lambda_{\pm} = -i\frac{\gamma}{2} \pm i|\epsilon| 
    \left( 1 - \frac{\theta^2}{2|\epsilon|^2} \right),\label{eigenvalues}
\end{equation}
which indicates that the eigenvalue associated with the amplified mode scales as $\theta^{2}$ at the PO threshold ($|\epsilon| = \frac{\gamma}{2}$). 

\subsection{Scattering properties}
To evaluate the system's response to the input, we calculate the Green's function. By transforming Eqs.~(\ref{EM1}) and (\ref{EM2}) into the frequency domain, we obtain
\begin{align}
&-i\omega a[\omega] = \big(-i\delta - \frac{\gamma}{2}\big)a[\omega] + \epsilon \{a[-\omega]\}^{\dagger} + \sqrt{\gamma_c} a_{\text{in}}[\omega] \nonumber\\&+ \sqrt{\gamma_0} b_{\text{in}}[\omega]- \sqrt{\kappa} \{c_{\text{in}}[-\omega]\}^{\dagger},\label{8}
\end{align}
\begin{align}
&-i\omega \{a[-\omega]\}^{\dagger} = \big(i\delta - \frac{\gamma}{2}\big)\{a[-\omega]\}^{\dagger} + \epsilon^* a[\omega] \nonumber\\&+ \sqrt{\gamma_c} \{a_{\text{in}}[-\omega]\}^{\dagger} + \sqrt{\gamma_0} \{b_{\text{in}}[-\omega]\}^{\dagger}- \sqrt{\kappa} c_{\text{in}}[\omega],\label{9}
\end{align}
where $\omega=\omega_d-\omega_p/2$ is the detuning between the frequency of driving field ($\omega_d$) and half of the pump frequency. Applying the substitution \(\omega \to -\omega\) to the above equations yields
\begin{align}
i\omega a[-\omega] &= \big(-i\delta - \frac{\gamma}{2}\big)a[-\omega] + \epsilon \{a[\omega]\}^{\dagger} + \sqrt{\gamma_c} a_{\text{in}}[-\omega]\nonumber\\&+ \sqrt{\gamma_0} b_{\text{in}}[-\omega]- \sqrt{\kappa} \{c_{\text{in}}[\omega]\}^{\dagger},\\
i\omega\{a[\omega]\}^{\dagger} &= \big(i\delta - \frac{\gamma}{2}\big)\{a[\omega]\}^{\dagger} + \epsilon^* a[-\omega] + \sqrt{\gamma_c} \{a_{\text{in}}[\omega]\}^{\dagger}\nonumber\\&+ \sqrt{\gamma_0} \{b_{\text{in}}[\omega]\}^{\dagger}- \sqrt{\kappa} c_{\text{in}}[-\omega].\label{11}
\end{align}
To study the dynamics in the quadrature basis, we introduce the quadratures in the frequency domain
\begin{align}
\hat{q}[\omega] &= \hat{a}[\omega] + \{a[\omega]\}^{\dagger},\\ \quad \hat{p}[\omega] &= -i\big(\hat{a}[\omega] - \{a[\omega]\}^{\dagger}\big),
\end{align}
which are Hermitian operators corresponding to measurable quantities. At $-\omega$ we have
\begin{align}
\hat{q}[-\omega] &= \hat{a}[-\omega] + \{a[-\omega]\}^{\dagger},\\
\quad \hat{p}[-\omega] &= -i\big(\hat{a}[-\omega] - \{a[-\omega]\}^{\dagger}\big).
\end{align}
Denoting the sequence as \(\{\hat{q}[\omega], \hat{q}[-\omega], \hat{p}[\omega], \hat{p}[-\omega]\}\), we combine Eqs.~(\ref{8})-(\ref{11}) and obtain
\begin{align}
-\omega q[\omega] =& -\delta q[\omega] -  \frac{\gamma}{2}  p[\omega] -\text{Re}(\epsilon)p[-\omega] +  \text{Im}(\epsilon) q[-\omega] \nonumber\\
&+ \sqrt{\gamma_{c}}  p_{\text{in}}[\omega] + \sqrt{\gamma_{0}}  p_{b,{\rm in}}[\omega] + \sqrt{\kappa}  p_{\text{c,in}}[-\omega], \label{F1}\\
-\omega q[-\omega] =& \delta q[-\omega] + \frac{\gamma}{2} p[-\omega] + \text{Re}(\epsilon)p[\omega] -  \text{Im}(\epsilon) q[\omega] \nonumber\\
&- \sqrt{\gamma_{c}}  p_{\text{in}}[-\omega] 
-\sqrt{\gamma_{0}}  p_{b,{\rm in}}[-\omega] - \sqrt{\kappa}  p_{\text{c,in}}[\omega], \label{F2}\\
-\omega p[\omega] =& -\delta p[\omega] + \frac{\gamma}{2} q[\omega] - \text{Re}(\epsilon) q[-\omega] - \text{Im}(\epsilon) p[-\omega]  \nonumber\\
&- \sqrt{\gamma_{c}}  q_{\text{in}}[\omega]- \sqrt{\gamma_{0}}  q_{b,{\rm in}}[\omega] + \sqrt{\kappa}  q_{\text{c,in}}[-\omega], \label{F3}\\
-\omega p[-\omega] =& \delta p[-\omega] - \frac{\gamma}{2} q[-\omega] + \text{Re}(\epsilon) q[\omega] + \text{Im}(\epsilon)p[\omega]  \nonumber\\
&+ \sqrt{\gamma_{c}}  q_{\text{in}}[-\omega] +\sqrt{\gamma_{0}}  q_{b,{\rm in}}[-\omega] - \sqrt{\kappa}  q_{\text{c,in}}[\omega]. \label{F4}
\end{align}
Reorganizing the above equations in the sequence Eqs.~(\ref{F3})(\ref{F4})(\ref{F1})(\ref{F2}) leads to
\begin{align}
&\frac{\gamma}{2} q[\omega] - \text{Re}(\epsilon) q[-\omega] + (\omega - \delta)p[\omega] - \text{Im}(\epsilon)p[-\omega] \nonumber\\&= \sqrt{\gamma_c} q_{\text{in}}[\omega] + \sqrt{\gamma_0} q_{b,\text{in}}[\omega] - \sqrt{\kappa} q_{c,\text{in}}[-\omega],\label{DF1}\\
&-\text{Re}(\epsilon) q[\omega] + \frac{\gamma}{2} q[-\omega] - \text{Im}(\epsilon)p[\omega] - (\omega + \delta)p[-\omega] \nonumber\\&= \sqrt{\gamma_c} q_{\text{in}}[-\omega] + \sqrt{\gamma_0} q_{b,\text{in}}[-\omega] - \sqrt{\kappa} q_{c,\text{in}}[\omega],\label{DF2}\\
&-(\omega - \delta) q[\omega] - \text{Im}(\epsilon) q[-\omega] + \frac{\gamma}{2} p[\omega] + \text{Re}(\epsilon) p[-\omega] \label{DF3}\\&= \sqrt{\gamma_c} p_{\text{in}}[\omega] + \sqrt{\gamma_0} p_{b,\text{in}}[\omega] + \sqrt{\kappa} p_{c,\text{in}}[-\omega],\nonumber\\
&-\text{Im}(\epsilon) q[\omega] + (\omega + \delta) q[-\omega] + \text{Re}(\epsilon) p[\omega] + \frac{\gamma}{2} p[-\omega] \nonumber\\&= \sqrt{\gamma_c} p_{\text{in}}[-\omega] + \sqrt{\gamma_0} p_{b,\text{in}}[-\omega] + \sqrt{\kappa} p_{c,\text{in}}[\omega].\label{DF4}
\end{align}
We thereby obtain the following equation of motion for quadratures in the frequency domain
\begin{align}
&-(\omega \bm{I} - \bm{M})\begin{pmatrix}
q[\omega] \\ q[-\omega] \\ p[\omega] \\ p[-\omega]
\end{pmatrix}
=\sqrt{\gamma_c}\bm{\Omega}
\begin{pmatrix}
q_{{\rm in}}[\omega] \\ q_{{\rm in}}[-\omega] \\ p_{{\rm in}}[\omega] \\ p_{{\rm in}}[-\omega]
\end{pmatrix}\nonumber\\
&+\sqrt{\gamma_0}\bm{\Omega}
\begin{pmatrix}
q_{b,{\rm in}}[\omega] \\ q_{b,{\rm in}}[-\omega] \\ p_{b,{\rm in}}[\omega] \\ p_{b,{\rm in}}[-\omega]
\end{pmatrix}
+\sqrt{\kappa}\bm{\Omega}^{'}
\begin{pmatrix}
q_{c,{\rm in}}[\omega] \\ q_{c,{\rm in}}[-\omega] \\ p_{c,{\rm in}}[\omega] \\ p_{c,{\rm in}}[-\omega]
\end{pmatrix},\label{EOM_qua}
\end{align}
where we define a non-Hermitian Hamiltonian under the quadrature basis
\begin{equation}
\bm{M} =
\begin{pmatrix}
\delta & -\text{Im}(\epsilon) & \frac{\gamma}{2} & \text{Re}(\epsilon) \\
\text{Im}(\epsilon) & -\delta & -\text{Re}(\epsilon) & -\frac{\gamma}{2} \\
-\frac{\gamma}{2} & \text{Re}(\epsilon) & \delta & \text{Im}(\epsilon) \\
-\text{Re}(\epsilon) & \frac{\gamma}{2} & -\text{Im}(\epsilon) & -\delta
\end{pmatrix},
\end{equation}
and two transformation matrices
\begin{align}
\bm{\Omega} &=
\begin{pmatrix}
0 & 0 &1 & 0 \\
0 & 0 & 0 & -1 \\
-1 & 0 & 0 & 0 \\
0 & 1 & 0 & 0
\end{pmatrix},\\
\bm{\Omega}^{'} &=
\begin{pmatrix}
0 & 0 &0 & 1 \\
0 & 0 & -1 & 0 \\
0 & 1 & 0 & 0 \\
-1 & 0 & 0 & 0
\end{pmatrix}.
\end{align}

We can further rewrite the equation of motion as of Eq.~(\ref{EOM_qua}) in the form 
\begin{align}
&\begin{pmatrix}
q[\omega] \\ q[-\omega] \\ p[\omega] \\ p[-\omega]
\end{pmatrix}
= \sqrt{\gamma_c}\bm{G}[\omega] 
\begin{pmatrix}
q_{\text{in}}[\omega] \\ q_{\text{in}}[-\omega] \\ p_{\text{in}}[\omega] \\ p_{\text{in}}[-\omega]
\end{pmatrix}
+\sqrt{\gamma_0}\bm{G}[\omega]
\begin{pmatrix}
q_{b,{\rm in}}[\omega] \\ q_{b,{\rm in}}[-\omega] \\ p_{b,{\rm in}}[\omega] \\ p_{b,{\rm in}}[-\omega]
\end{pmatrix}\nonumber\\
&+\sqrt{\kappa}\bm{G}^{'}[\omega]
\begin{pmatrix}
q_{c,{\rm in}}[\omega] \\ q_{c,{\rm in}}[-\omega] \\ p_{c,{\rm in}}[\omega] \\ p_{c,{\rm in}}[-\omega]
\end{pmatrix},
\end{align}
with the main and secondary Green's functions defined as
\begin{align}
\bm{G}[\omega] &= -(\omega \bm{I} - \bm{M})^{-1}\bm{\Omega} =-\bm{\Omega} (\omega \bm{I} - \bm{M})^{-1} \nonumber\\&= 
\begin{pmatrix}
\frac{\gamma}{2} & -\text{Re}(\epsilon) & \omega - \delta & -\text{Im}(\epsilon) \\
-\text{Re}(\epsilon) & \frac{\gamma}{2} & -\text{Im}(\epsilon) & -(\omega + \delta) \\
-(\omega - \delta) & -\text{Im}(\epsilon) & \frac{\gamma}{2} & \text{Re}(\epsilon) \\
-\text{Im}(\epsilon) & (\omega + \delta) & \text{Re}(\epsilon) & \frac{\gamma}{2}
\end{pmatrix}^{-1}.\\
\bm{G}^{'}[\omega] &= -(\omega \bm{I} - \bm{M})^{-1}\bm{\Omega}^{'} =-\bm{\Omega}^{'} (\omega \bm{I} - \bm{M})^{-1}.
\end{align}
Utilizing the input-output relation
\begin{equation}
\begin{pmatrix}
q_{\text{out}}[\omega] \\ q_{\text{out}}[-\omega] \\ p_{\text{out}}[\omega] \\ p_{\text{out}}[-\omega]
\end{pmatrix}
= \begin{pmatrix}
q_{\text{in}}[\omega] \\ q_{\text{in}}[-\omega] \\ p_{\text{in}}[\omega] \\ p_{\text{in}}[-\omega]
\end{pmatrix}
- \sqrt{\gamma_c}
\begin{pmatrix}
q[\omega] \\ q[-\omega] \\ p[\omega] \\ p[-\omega]
\end{pmatrix},
\end{equation}
we obtain
\begin{align}
&\begin{pmatrix}
q_{\text{out}}[\omega] \\ q_{\text{out}}[-\omega] \\ p_{\text{out}}[\omega] \\ p_{\text{out}}[-\omega]
\end{pmatrix}
= \bm{S}[\omega] \begin{pmatrix}
q_{\text{in}}[\omega] \\ q_{\text{in}}[-\omega] \\ p_{\text{in}}[\omega] \\ p_{\text{in}}[-\omega]
\end{pmatrix}
+ \bm{L}[\omega]
\begin{pmatrix}
q_{b,{\rm in}}[\omega] \\ q_{b,{\rm in}}[-\omega] \\ p_{b,{\rm in}}[\omega] \\ p_{b,{\rm in}}[-\omega]
\end{pmatrix}\nonumber\\
&+\bm{K}[\omega]
\begin{pmatrix}
q_{c,{\rm in}}[\omega] \\ q_{c,{\rm in}}[-\omega] \\ p_{c,{\rm in}}[\omega] \\ p_{c,{\rm in}}[-\omega]
\end{pmatrix},
\end{align}
with the scattering matrix defined as
\begin{equation}
\bm{S}[\omega] = \bm{I} - \gamma_c \bm{G}[\omega],
\end{equation}
and the scattering matrices for the intrinsic and amplification noise given by
\begin{equation}
\bm{L}[\omega] = -\sqrt{\gamma_0 \gamma_c} \bm{G}[\omega],
\end{equation}
\begin{equation}
\bm{K}[\omega] = -\sqrt{\kappa \gamma_c} \bm{G}^{'}[\omega].
\end{equation}

\subsection{Quantum Fisher information and Cramér-Rao bound}
We consider the sensing scheme that a perturbation is applied on the resonance frequency, such that the resonance-pump detuning is shifted $\delta\rightarrow\delta+\theta$. We assume that the probe-pump detuning $\omega$ is not affected by the perturbation. Our goal is to estimate the unknown parameter $\theta$ by performing measurements on the output field and to infer the value of perturbation $\theta$ based on the measurement results. There exists a smallest achievable deviation, given by the quantum Fisher Information via the Cramér-Rao Bound
\begin{equation}
    \Delta \theta \geq \frac{1}{\sqrt{N_m} \sqrt{\mathcal{I}(\theta)}},
\end{equation}
where $N_m$ is the number of rounds of measurement, and $\mathcal{I}(\theta)$ is the quantum Fisher Information (QFI). In the following, we use QFI to estimate the lower bound of the sensing precision for a single-round measurement.

For a Gaussian state with a mean amplitude $\bm{\mu}_\theta$ and a covariance matrix $\bm{V}_\theta$ in the quadrature basis, the QFI for estimating the parameter $\theta$ is given as \cite{zhang2019quantum}
\begin{equation}
    \mathcal{I}(\theta) = 
    \left( \frac{d \bm{\mu}_\theta}{d\theta} \right)^T \bm{V}_\theta^{-1} 
    \left( \frac{d \bm{\mu}_\theta}{d\theta} \right) 
    + \frac{1}{2} \text{Tr} \left( \bm{\Phi}_\theta \frac{d \bm{V}_\theta}{d\theta} \right),
\end{equation}
where
\begin{align}
    \mathcal{I}_\mu(\theta) &= \left( \frac{d \bm{\mu}_\theta}{d\theta} \right)^T \bm{V}_\theta^{-1} 
    \left( \frac{d \bm{\mu}_\theta}{d\theta} \right), \quad \\
    \mathcal{I}_V(\theta) &= \frac{1}{2} \text{Tr} \left( \bm{\Phi}_\theta \frac{d \bm{V}_\theta}{d\theta} \right).
\end{align}
The matrix $\bm{\Phi}_\theta$ is implicitly determined by:
\begin{equation}
    \frac{d \bm{V}_\theta}{d\theta} = \bm{V}_\theta \bm{\Phi}_\theta \bm{V}_\theta - \bm{\tilde{\Omega}} \bm{\Phi}_\theta \bm{\tilde{\Omega}}^T,
\end{equation}
where $\bm{\tilde{\Omega}}$ is the fundamental symplectic matrix satisfying $\bm{\tilde{\Omega}}^2 = -\bm{I}$. For noisy Gaussian state, $\mathcal{I}_{V}(\theta)$ is reduced to a simplified form \cite{jiang2014quantum, zhang2019quantum}
\begin{align}
\mathcal{I}_{V}(\theta) &= \frac{1}{2} \, \mathrm{Tr} \left( 
\bm{V}_\theta^{-1} \frac{d \bm{V}_\theta}{d\theta} 
\bm{V}_\theta^{-1} \frac{d \bm{V}_\theta}{d\theta} 
\right).\label{Iv}
\end{align}

To evaluate QFI we calculate the derivative of the mean of output
\begin{align}
    \frac{d\bm{\mu}_{\theta}}{d\theta} &= -\gamma_c \frac{d\bm{G}_{\theta}}{d\theta} \bm{\mu}_{\rm in} = -\gamma_c (-\bm{G}_{\theta}) \frac{d[({{\omega} \bm{I}-\bm{M}_{\theta}})\bm{\Omega}]}{d\theta} \bm{G}_{\theta} \bm{\mu}_{\rm in}\nonumber\\ &= -\gamma_c \bm{G}_{\theta} {\Pi} \bm{\Omega} \bm{G}_{\theta} \bm{\mu}_{\rm in} = -\gamma_c \bm{G}_{\theta} { \bm{\tilde{\Omega}}} \bm{G}_{\theta} \bm{\mu}_{\rm in}.
\end{align}
where
\begin{equation}
    \bm{\Pi} = 
\begin{pmatrix}
1 & 0 &0 & 0 \\
0 & -1 & 0 & 0 \\
0 & 0 & 1 & 0 \\
0 & 0 & 0 & -1
\end{pmatrix},
\end{equation}
and
\begin{equation}
\bm{\tilde{\Omega}} =
\begin{pmatrix}
0 & 0 &1 & 0 \\
0 & 0 & 0 & 1 \\
-1 & 0 & 0 & 0 \\
0 & -1 & 0 & 0
\end{pmatrix}.
\end{equation}
Further, the covariance matrix can be expressed as 
\begin{align}    \bm{V}_{\theta}&=\bm{S}\bm{V}_{\rm in}\bm{S}^{T}+\bm{L}\bm{V}_{\rm in}^{'}\bm{L}^{T}+\bm{K}\bm{V}_{\rm in}^{''}\bm{K}^{T}\nonumber\\&=(\bm{I}-\gamma_c \bm{G}_{\theta})\bm{V}_{\rm in}(\bm{I}- \gamma_c \bm{G}_{\theta})^{T}+\gamma_0 \gamma_c \bm{G}_{\theta} \bm{V}_{\rm in}^{'}\bm{G}_{\theta}^{T}\nonumber\\
&+\kappa \gamma_c \bm{G}_{\theta}^{'}\bm{V}_{\rm in}^{''}\bm{G}_{\theta}^{'T},
\end{align}
Therefore we have
\begin{align}
    \mathcal{I}_{\mu}(\theta) &= \gamma_c^2 \bm{\mu}_{\text{in}}^T \bm{G}_\theta^T \bm{\tilde{\Omega}}^T\bm{G}_\theta^T \nonumber\\
    &\times \left[ 
        (\bm{I} - \gamma_c \bm{G}_\theta)^T \bm{V}_{\text{in}} (\bm{I} - \gamma_c \bm{G}_\theta)
        + \gamma_0 \gamma_c \bm{G}_{\theta} \bm{V}_{\text{in}}^{'} \bm{G}_{\theta} \right. \nonumber\\
    &\qquad \left. 
        + \kappa \gamma_c \bm{G}_{\theta}^{'} \bm{V}_{\text{in}}^{''} \bm{G}_{\theta}^{'}
    \right]^{-1} \times \bm{G}_\theta \bm{\tilde{\Omega}} \bm{G}_\theta \bm{\mu}_{\text{in}}.
\end{align}
On the other hand, 
\begin{align}
\frac{d \bm{V}_\theta}{d\theta}
={}&-\gamma_c\Big(\, \bm{G}_\theta \tilde{\bm{\Omega}} \bm{G}_\theta \bm{V}_{\text{in}} \bm{S}^{T}
+\bm{S} \bm{V}_{\text{in}}\bm{G}^{T}_\theta \tilde{\bm{\Omega}}^{T} \bm{G}_\theta^{T}
      \Big)  \nonumber\\
& - \sqrt{\gamma_0\gamma_c}\Big(\, \bm{G}_\theta \tilde{\bm{\Omega}} \bm{G}_\theta \bm{V}_{\text{in}}' \bm{L}^{T}
      + \bm{L} \bm{V}_{\text{in}}' \bm{G}^{T}_\theta \tilde{\bm{\Omega}}^{T}  \bm{G}^{T}_\theta \Big) \nonumber\\
& - \sqrt{\kappa\,\gamma_c}\Big(\, \bm{G}'_\theta \tilde{\bm{\Omega}}' \bm{G}'_\theta \bm{V}_{\text{in}}'' \bm{K}^{T}
      + \bm{K} \bm{V}_{\text{in}}'' \bm{G}'^{T}_\theta \tilde{\bm{\Omega}}'^{T} \bm{G}'^{T}_\theta \Big),
\end{align}
which allows us to compute $\mathcal{I}_{V}(\theta)$ using Eq.~(\ref{Iv}).

Additionally, for large enough input, $\mathcal{I}_{\mu}(\theta)$ dominates, which determines the scaling of QFI with perturbation. It can be seen that the lower bound of detection can be amplified by large $\bm{G}_{\theta}$. While the SNR of the output signal is quantified by $\bm{\mu}_{\theta}^T \bm{V}_{\theta}^{-1} \bm{\mu}_{\theta}$, in the process that the perturbation is transduced to the output, the system's response is characterized by a transduction function $\frac{d\bm{\mu}_{\theta}}{d\theta}$ which scales linearly with $\bm{G}_{\theta}$. Thus, amplifying the transduction function through $\bm{G}_{\theta}$ leads to enhanced SNR and sensitivity.

As a result, for a single-mode sensor operating at the PO threshold, the lower bound of the sensing precision scales with the perturbation as
\begin{equation}
    \delta \theta \geq \frac{1}{\sqrt{\mathcal{I}(\theta)}} \sim \theta^{2}.
\end{equation}

\section{Two-mode bosonic sensor}
\subsection{Quantum dynamics}
For the two-mode scheme, the Hamiltonian of the system is given by:
\begin{align}
H &= \hbar \delta_1 a_1^\dagger a_1 + \hbar \delta_2 a_2^\dagger a_2 + \hbar g (a_1^\dagger a_2 + \text{H.c.}) \nonumber\\
&+ \frac{i \hbar}{2} (\epsilon_1 a_1^{\dagger 2} - \epsilon_1^* a_1^2) + \frac{i \hbar}{2} (\epsilon_2 a_2^{\dagger 2} - \epsilon_2^* a_2^2).
\end{align}
Following the analysis for the single-mode scheme, we get the equations of motion under probes $a_{in,1}$ and $a_{in,2}$
\begin{align}
\frac{d a_1(t)}{dt} &= (-i \delta_1 - \frac{\gamma_{1}}{2}) a_1 - i g a_2 + \epsilon_1 a_1^\dagger + \sqrt{\gamma_{c1}} a_{\text{in,1}} \nonumber\\
&+ \sqrt{\gamma_{01}} b_{\text{in,1}} - \sqrt{\kappa_{1}} c_{\text{in,1}}^\dagger,\\
\frac{d a_2(t)}{dt} &= (-i \delta_2 - \frac{\gamma_{2}}{2}) a_2 - i g a_1 + \epsilon_2 a_2^\dagger + \sqrt{\gamma_{c2}} a_{\text{in,2}} \nonumber\\
&+ \sqrt{\gamma_{02}} b_{\text{in,2}} - \sqrt{\kappa_{2}} c_{\text{in,2}}^\dagger,\\
\frac{d a_1^\dagger(t)}{dt} &= (i \delta_1 - \frac{\gamma_{1}}{2}) a_1^\dagger + i g a_2^\dagger + \epsilon_1^* a_1 + \sqrt{\gamma_{c1}} a_{\text{in,1}}^\dagger \nonumber\\
&+ \sqrt{\gamma_{01}} b_{\text{in,1}}^\dagger - \sqrt{\kappa_{1}} c_{\text{in,1}},\\
\frac{d a_2^\dagger(t)}{dt} &= (i \delta_2 - \frac{\gamma_{2}}{2}) a_2^\dagger + i g a_1^\dagger + \epsilon_2^* a_2 + \sqrt{\gamma_{c2}} a_{\text{in,2}}^\dagger \nonumber\\
&+ \sqrt{\gamma_{02}} b_{\text{in,2}}^\dagger - \sqrt{\kappa_{2}} c_{\text{in,2}}.
\end{align}
Here $g$ is the coupling rate between modes $a_1$ and $a_2$. We denote the loss rates in the $j$-th resonators as $\gamma_{0j}$, the out-coupling loss rates as $\gamma_{cj}$, and the amplification rates as $\kappa_j$. The total loss rates are represented by
$\gamma_j = \gamma_{0i} + \gamma_{ci}-\kappa_i$.

By transforming the equation of motion into the frequency domain, we get
\begin{align}
-i \omega a_1[\omega] &= \left(-i \delta_1 - \frac{\gamma_1}{2}\right) a_1[\omega] - i g a_2[\omega] + \epsilon_1 \{a_1[-\omega]\}^\dagger \nonumber\\
&+ \sqrt{\gamma_{c1}} a_{\text{in}1}[\omega] + \sqrt{\gamma_{02}} b_{\text{in}1}[\omega] - \sqrt{\kappa_{1}} \{c_{\text{in}1}[-\omega]\}^\dagger,\\
-i \omega a_2[\omega] &= \left(-i \delta_2 - \frac{\gamma_2}{2}\right) a_2[\omega] - i g a_1[\omega] + \epsilon_2 \{a_2[-\omega]\}^\dagger \nonumber\\
&+ \sqrt{\gamma_{c2}} a_{\text{in}2}[\omega]+ \sqrt{\gamma_{02}} b_{\text{in}2}[\omega] - \sqrt{\kappa_{2}} \{c_{\text{in}2}[-\omega]\}^\dagger.
\end{align}
Converting to quadrature basis yields
\begin{align}
    -\omega q_1[\omega] &= -\delta_1 q_1[\omega] -g q_2[\omega] -  \frac{\gamma_1}{2}  p_1[\omega] -\text{Re}(\epsilon_1)p_1[-\omega] \nonumber\\
    &+  \text{Im}(\epsilon_1) q_1[-\omega] + \sqrt{\gamma_{c1}}  p_{\text{in1}}[\omega] + \sqrt{\gamma_{01}}  p_{\text{b,in1}}[\omega] \nonumber\\
    &+ \sqrt{\kappa_1}  p_{\text{c,in1}}[-\omega],\\
    -\omega q_1[-\omega] &= \delta_1 q_1[-\omega] +g q_2[-\omega] + \frac{\gamma_1}{2} p_1[-\omega] + \text{Re}(\epsilon_1)p_1[\omega] \nonumber\\
    &-\text{Im}(\epsilon_1) q_1[\omega] - \sqrt{\gamma_{c1}}  p_{\text{in1}}[-\omega] - \sqrt{\gamma_{01}}  p_{\text{b,in1}}[-\omega] \nonumber\\
    &- \sqrt{\kappa_1}  p_{\text{c,in1}}[\omega],\\ 
    -\omega p_1[\omega] &= -\delta_1 p_1[\omega] -g p_2[\omega]+ \frac{\gamma_1}{2} q_1[\omega] - \text{Re}(\epsilon_1) q_1[-\omega] \nonumber \\ 
    &- \text{Im}(\epsilon_1) p_1[-\omega]- \sqrt{\gamma_{c1}}  q_{\text{in1}}[\omega] - \sqrt{\gamma_{01}}  q_{\text{b,in1}}[\omega] \nonumber \\ 
    &+ \sqrt{\kappa_1}  q_{\text{c,in1}}[-\omega],\\
    -\omega p_1[-\omega] &= \delta_1 p_1[-\omega] +g p_2[-\omega]- \frac{\gamma_1}{2} q_1[-\omega] + \text{Re}(\epsilon_1) q_1[\omega] \nonumber\\ &+ \text{Im}(\epsilon_1)p_1[\omega] + \sqrt{\gamma_{c1}}  q_{\text{in1}}[-\omega] + \sqrt{\gamma_{01}}  q_{\text{b,in1}}[-\omega] \nonumber\\ &- \sqrt{\kappa_1}  q_{\text{c,in1}}[\omega]. 
\end{align}
Moreover, we can interchange subscript 1 and 2 to derive the other four equations of motion in the frequency domain for the quadratures of mode 2. 
The quadratures of the full system can be grouped in the sequence:
\begin{align}
\textbf{a}[\omega] &= \{ \hat{q}_1[\omega], \hat{q}_1[-\omega], \hat{p}_1[\omega], \hat{p}_1[-\omega], \hat{q}_2[\omega], \hat{q}_2[-\omega],\nonumber\\ &\hat{p}_2[\omega], \hat{p}_2[-\omega] \}^T.
\end{align}
The 8-by-8 non-Hermitian Hamiltonian under the quadrature basis is then given by 
\begin{equation}
    \bm{M} = \begin{pmatrix}
    \bm{M}_1 & \bm{M}_{12} \\
    \bm{M}_{12} & \bm{M}_2
    \end{pmatrix},\end{equation}
where
\begin{equation}
\bm{M}_{j} = 
\begin{pmatrix}
\delta_{j} & -\text{Im}(\epsilon_{j}) & \frac{\gamma_{j}}{2} & \text{Re}(\epsilon_{j}) \\
\text{Im}(\epsilon_{j}) & -\delta_{j} & -\text{Re}(\epsilon_{j}) & -\frac{\gamma_{j}}{2} \\
-\frac{\gamma_{j}}{2} & \text{Re}(\epsilon_{j}) & \delta_{j} & \text{Im}(\epsilon_{j})\\
-\text{Re}(\epsilon_{j}) & \frac{\gamma_{j}}{2} & -\text{Im}(\epsilon_{j}) & -\delta_{j} \\
\end{pmatrix},\label{Mj}
\end{equation}
for $j=1,2$ and $\bm{M}_{12} = \mathrm{diag}\{g, -g, g, -g\}$. We can thereby obtain the dynamic equation
\begin{equation}
-\omega \mathbf{a}[\omega] =
-\bm{M} \mathbf{a}[\omega] + \bm{\Omega}_a \mathbf{a}_{\text{in}}[\omega] 
+ \bm{\Omega}_b \mathbf{b}_{\text{in}}[\omega] + \bm{\Omega}_c \mathbf{c}_{\text{in}}[\omega],
\end{equation}
and the input-output relation:
\begin{equation}
\mathbf{a_{\text{out}}}[\omega] = \mathbf{a_{\text{in}}}[\omega] - \bm{K}_{\rm ex}\mathbf{a},
\end{equation}
where
\begin{align}
\bm{\Omega}_a &= (\bm{\Omega} \otimes \bm{I}_{2}) \bm{K}_{\rm ex},\\
\bm{\Omega}_b &= (\bm{\Omega} \otimes \bm{I}_{2}) \bm{K}_{i},\\
\bm{\Omega}_c &= (\bm{\Omega}^{'} \otimes \bm{I}_{2}) \bm{K}_{g},
\end{align}
and
\begin{align}
\bm{K}_{\rm ex} &= \begin{pmatrix}
\sqrt{\gamma_{c1}}\bm{I}_{4} & 0\\ 0 & \sqrt{\gamma_{c2}}\bm{I}_{4}
\end{pmatrix},\\
\bm{K}_{i} &= \begin{pmatrix}
\sqrt{\gamma_{i1}}\bm{I}_{4} & 0\\ 0 & \sqrt{\gamma_{i2}}\bm{I}_{4}
\end{pmatrix},\\
\bm{K}_{g} &= \begin{pmatrix}
\sqrt{\kappa_{1}}\bm{I}_{4} & 0\\ 0 & \sqrt{\kappa_{2}}\bm{I}_{4}
\end{pmatrix}.
\end{align}

\subsection{Scattering properties}
Similar to the single-mode scheme discussed in the previous section and the two-mode scheme studied in \cite{zhang2019quantum}, we define the Green's functions \( \bm{G}[\omega] \) and \( \bm{G}^{'}[\omega] \) as:
\begin{equation}
    \bm{G}[\omega] = -(\omega \bm{I} - \bm{M})^{-1} (\bm{I}_{2} \otimes \bm{\Omega}),
\end{equation}
\begin{equation}
    \bm{G}^{'}[\omega] = -(\omega \bm{I} - \bm{M})^{-1} (\bm{I}_{2} \otimes \bm{\Omega^{'}}).
\end{equation}
Then we get
\begin{equation}
    \mathbf{a_{\text{out}}}[\omega] = \bm{S}[\omega] \mathbf{a_{\text{in}}}[\omega] + \bm{L}[\omega] \mathbf{b_{\text{in}}}[\omega] + \bm{K}[\omega] \mathbf{c_{\text{in}}[\omega]},
\end{equation}
where
\begin{equation}
    \bm{S}[\omega] = \bm{I}- \bm{K}_{\rm ex} \bm{G}_{\theta}[\omega] \bm{K}_{\rm ex},
\end{equation}\begin{equation}    
    \bm{L}[\omega] = -\bm{K}_{\rm ex} \bm{G}_{\theta}[\omega] \bm{K}_i,
\end{equation}\begin{equation}
    \bm{K}[\omega] = -\bm{K}_{\rm ex} \bm{G}_{\theta}^{'}[\omega]\bm{K}_g.
\end{equation}
The determinant of the scattering matrix \( \bm{S}[\omega] \) scales as a higher power of \( \theta \). Below, we will study the scaling of the QFI with respect to $\theta$ when the system is operating at an EP.

\section{Exceptional point enhanced sensing}
\subsection{Precision scaling of EP sensors at the PO threshold}
We first review the case of EP sensing at the lasing threshold discussed in \cite{zhang2019quantum}. For $n$-th order EP at the lasing threshold, we have $M_{\rm EP}^n=0$. For the two mode case with a 2nd-order EP, at the lasing threshold without the effect of phase-sensitive amplification, the Hamiltonian under the eigenbasis reduces to a Jordan normal form
\begin{equation}
    \bm{M}_{\rm EP} = 
    \begin{bmatrix}
        0 & 1\\
        0 & 0
    \end{bmatrix}.
\end{equation}
In general, for an nth-order EP, the Green's function is expressed as
\begin{align}
    \bm{G}_\theta &= - \left( \theta \bm{I} - \bm{M}_{\rm EP} \right)^{-1} \bm{\Omega}
    = -\theta^{-1} \left( \bm{I} - \theta^{-1} \bm{M}_{\rm EP} \right)^{-1} \bm{\Omega}\nonumber\\
    &= -\theta^{-1} \bm{I} \bm{\Omega} - \theta^{-2} \bm{M}_{\rm EP} \bm{\Omega}
    - \dots - \theta^{-n} \bm{M}_{\rm EP}^{n-1} \bm{\Omega},
\end{align}
where we have applied Taylor expansion and utilized the condition $\bm{M}_{\rm EP}^n=0$. The leading order of $\bm{G}_{\theta}$ scaling as $\theta^{-n}$ leads to the Cramér-Rao bound as reported in \cite{zhang2019quantum}
\begin{equation}
    \delta\theta_{\text{CRB}} \sim \frac{1}{\sqrt{\mathcal{I}_\theta}} \sim \theta^n.
\end{equation}

In the case of single-mode squeezing near the PO threshold at $\delta=0$, two eigenvalues of the Hamiltonian scale as $\theta^2$, while the other two have a constant leading order. In the two-mode case, we can further tune the parameters to steer the coupled modes to reach a non-Hermitian degeneracy and the PO threshold simultaneously. When this condition is satisfied, four eigenvalues become pairwise degenerate and each scales as $\theta^2$, while the remaining four eigenvalues have a constant leading-order term. As a consequence, the determinant has a $\theta^8$ scaling. To derive the corresponding parameter condition from the above analysis, we examine the determinant of the Hamiltonian under the condition $\delta_{1,2}=0$
\begin{align}
\Delta &= 
\frac{1}{64} \left[ 
(4g^2 + \gamma_1 \gamma_2)^2 
+ 4(-8g^2 + \gamma_1^2 + \gamma_2^2)\theta^2 \right. \nonumber \\
&\quad + 16\theta^4  
- 4(4g^2 \epsilon_1 + \gamma_1^2 \epsilon_2 + 4 \epsilon_2 \theta^2) \epsilon_2^* \nonumber \\
&\quad \left. - 4 \epsilon_1^* (\gamma_2^2 \epsilon_1 + 4g^2 \epsilon_2 + 4 \epsilon_1 \theta^2 
- 4 \epsilon_1 \epsilon_2 \epsilon_2^*) 
\right]^2.
\end{align}
As can be seen, the coefficients of $\theta^2$ and $\theta^0$ in the square bracket must vanish to ensure that the determinant scales as $\theta^8$. Therefore the necessary conditions for reaching a second-order EP and the PO threshold under zero pump detuning ($\delta_{1,2}=0$) are 
\begin{align}
&|\epsilon_1|^2 + |\epsilon_2|^2 
= \frac{\gamma_1^2 + \gamma_2^2}{4} - 2g^2,
\end{align}
\begin{align}
-(g^2 + \frac{\gamma_1 \gamma_2}{4} )^2
&= |\epsilon_1|^2 |\epsilon_2|^2 
- \frac{\gamma_1^2 |\epsilon_2|^2 + \gamma_2^2 |\epsilon_1|^2}{4} \nonumber\\
&- 2g^2 \left[ \text{Re}(\epsilon_1)\text{Re}(\epsilon_2) + \text{Im}(\epsilon_1)\text{Im}(\epsilon_2) \right]. 
\end{align}
In spite of the fact that these equations guarantee the PO threshold condition for both modes, the existence of an EP must still be verified by checking the degeneracy of the eigenvalues and eigenstates. Notably, in absence of inter-modal coupling, i.e., $g=0$, it is still possible for the determinant to scale as $\theta^{-m}$ with $m>4$, indicating that two pairs of degenerate eigenvalues may have $\theta^2$ scaling without coupling. However, such trivial degeneracy is simply achieved by adding additional uncoupled modes with squeezing and does not result in improvement in the precision scaling. Mathematically, this effect can be understood by examining a Hamiltonian composed of submatrices that share identical eigenvalues. When inverting the Hamiltonian matrix, the determinant appears only once in the denominator. This is therefore a trivial degeneracy (often known as a diabolic point).

Based on the conditions to reach an EP and PO threshold simultaneously, we analyze the sensitivity enhancement under the joint EP and squeezing effects. As noted, the desired Hamiltonian under the eigenbasis takes the Jordan normal form
\begin{equation}
    \bm{\tilde{M}} = \begin{bmatrix}
        f_1 & 1 & 0 & 0 & 0 & 0 & 0 & 0 \\
        0 & f_1 & 0 & 0 & 0 & 0 & 0 & 0 \\
        0 & 0 & -f_1 & 1 & 0 & 0 & 0 & 0 \\
        0 & 0 & 0 & -f_1 & 0 & 0 & 0 & 0 \\
        0 & 0 & 0 & 0 & f_2 & 1 & 0 & 0 \\
        0 & 0 & 0 & 0 & 0 & f_2 & 0 & 0 \\
        0 & 0 & 0 & 0 & 0 & 0 & -f_2 & 1 \\
        0 & 0 & 0 & 0 & 0 & 0 & 0 & -f_2
    \end{bmatrix},
\end{equation}
where $f_1=a\theta^2$ and $f_2=b+a\theta^2$. Let $\bm{P}$ be the matrix that rotates the quadrature-basis Hamiltonian $\bm{M}$ to such a Jordan normal form. At $\omega=0$, the Green's function \( \bm{G}_{\theta} \) is given by:
\begin{align}
\bm{G}_{\theta} &= -(\omega \bm{I} - \bm{M})^{-1} (\bm{I}_{2} \otimes \bm{\Omega})\nonumber\\ &=
-\left(- \bm{P} \bm{\tilde{M}} {\bm{P}}^{-1} \right)^{-1} (\bm{I}_{2} \otimes \bm{\Omega}) \nonumber\\ & \approx \bm{P} \text{diag}\{
(a\theta^2 \bm{I}_2 + \bm{M}_{\text{EP}})^{-1}, (-a\theta^2 \bm{I}_2 + \bm{M}_{\text{EP}})^{-1},\nonumber\\ &[(b + a\theta^2)\bm{I}_2 + \bm{M}_{\rm EP}]^{-1}, [(-b - a\theta^2)\bm{I}_2 + \bm{M}_{\rm EP}]^{-1} \} \nonumber\\
&\times\bm{P}^{-1}(\bm{I}_{2} \otimes \bm{\Omega}).
\end{align}
Using the series expansion with $\bm{M}_{\text{EP}}^2=0$, we get
\begin{align}
(a\theta^2 \bm{I}_2 + \bm{M}_{\text{EP}})^{-1} &=  a^{-1} \theta^{-2} \left( \bm{I}_2 + a^{-1} \theta^{-2} \bm{M}_{\text{EP}} \right)^{-1} \nonumber\\
&=  a^{-1} \theta^{-2} \left( \bm{I}_2 - a^{-1} \theta^{-2} \bm{M}_{\text{EP}} \right).
\end{align}
Therefore,
\begin{align}
\bm{G}_{\theta}\approx & \bm{P} \left( \begin{array}{ccc} -a^{-2} \theta^{-4} \bm{M}_{\text{EP}} & \bm{0} & \bm{0} \\
\bm{0} & -a^{-2} \theta^{-4}\bm{M}_{\text{EP}}  & \bm{0}\\
\bm{0} & \bm{0} & \bm{F}_2
\end{array} \right) \bm{P}^{-1}\nonumber\\
&\times(\bm{I}_{2} \otimes \bm{\Omega}),
\end{align}
where $\bm{F}_2=\text{diag}\{[(b + a\theta^2) \bm{I}_2 + \bm{M}_{\text{EP}}]^{-1}, [(-b - a\theta^2) \bm{I}_2 + \bm{M}_{\text{EP}}]^{-1}\}$. It is noted that the leading order of each element of $\bm{P}$ is constant and nonzero. As a result, the Cramér-Rao bound scales with $\theta$ as
\begin{equation}
    \delta\theta_{\text{CRB}} \sim \frac{1}{\sqrt{\mathcal{I}_\theta}} \sim \theta^{4}.
\end{equation}

Now we generalize our result to an nth-order EP sensor at the PO threshold. Using the series expansion, we get
\begin{align}
(a\theta^2 \bm{I}_n + \bm{M}_{\text{EP},n})^{-1} &= a^{-1} \theta^{-2} \left( \bm{I}_n + a^{-1} \theta^{-2} \bm{M}_{\text{EP},n} \right)^{-1} \nonumber\\
&= a^{-1} \theta^{-2}  \left( \bm{I}_n - a^{-1} \theta^{-2} \bm{M}_{\text{EP},n} \right.\quad\nonumber\\ &+ \dots  
\left. +\, (-a)^{-(n-1)} \theta^{-2(n-1)} \bm{M}_{\text{EP},n}^{n-1} \right).
\end{align}
Since \( \bm{M}_{\text{EP},n}^n = 0 \), the expansion series terminates at the
$n$th term, with all higher-order terms vanishing. Therefore,
\begin{align}
\bm{G}_{\theta}\approx & \bm{P} \left( \begin{array}{ccc} -(-a)^{-n} \theta^{-2n} \bm{M}_{\text{EP},n}^{n-1} & \bm{0} & \bm{0} \\
\bm{0} & -a^{-n} \theta^{-2n}\bm{M}_{\text{EP},n}^{n-1}  & \bm{0}\\
\bm{0} & \bm{0} & \bm{F}_n
\end{array} \right)\nonumber\\
& \times\bm{P}^{-1}(\bm{I}_{2} \otimes \bm{\Omega}),\label{G_scaling}
\end{align}
where $\bm{F}_n=\text{diag}\{[(b + a\theta^2) \bm{I}_n + \bm{M}_{\text{EP},n}]^{-1}, [(-b - a\theta^2) \bm{I}_n + \bm{M}_{\text{EP},n}]^{-1}\}$. The Cramér-Rao bound therefore scales with $\theta$ as
\begin{equation}
    \delta\theta_{\text{CRB}} \sim \frac{1}{\sqrt{\mathcal{I}_\theta}} \sim \theta^{2n}.
\end{equation}

\subsection{Sensing at a third-order EP}
\begin{figure}[!htb] 	\centering\includegraphics[width=0.46\textwidth]{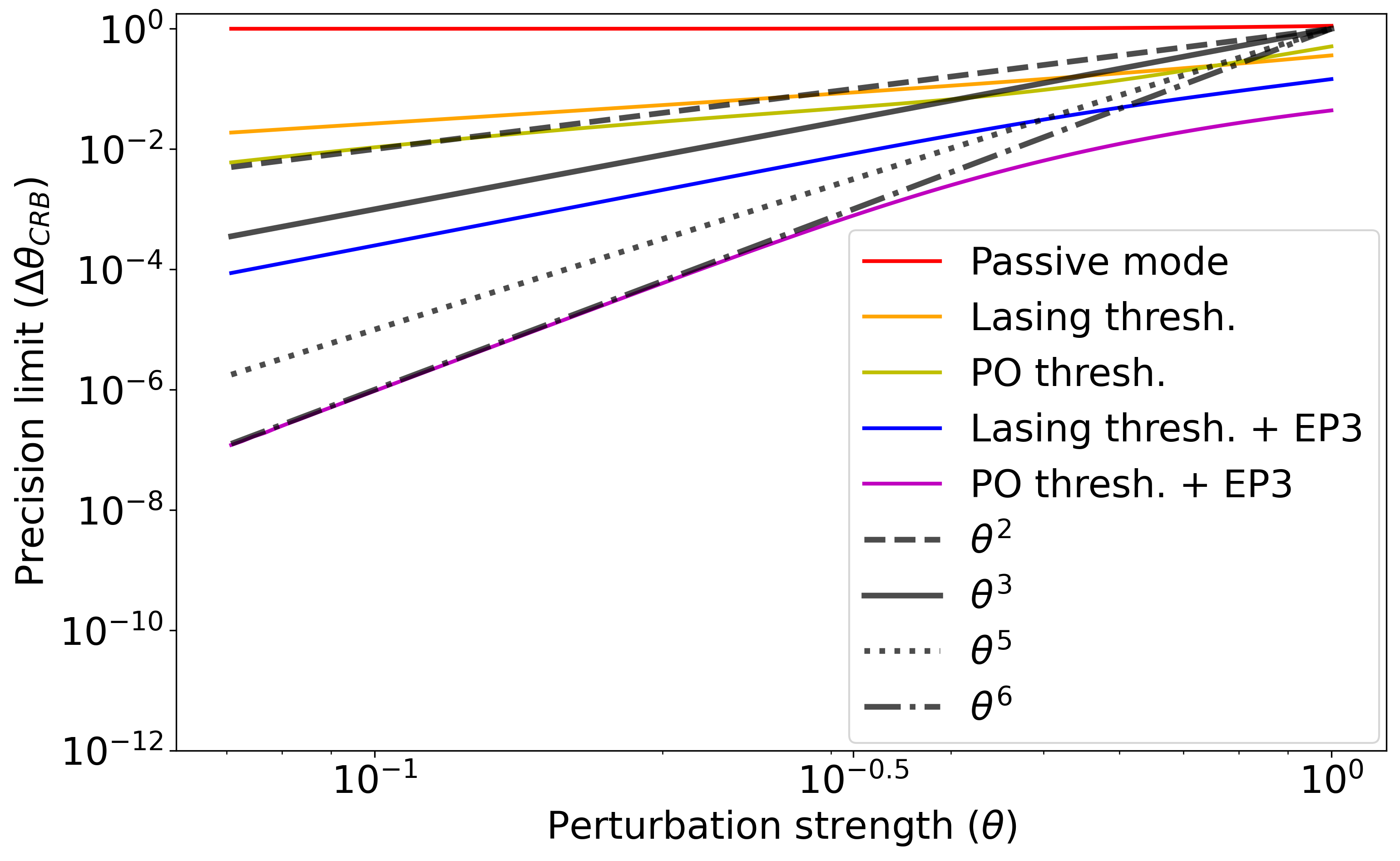}
	\caption{Precision limit (Cramér-Rao bound) as a function of the perturbation strength for a third-order EP (EP3) sensors operating at the lasing threshold and parametric-oscillation (PO) threshold, in comparison with other cases.}
\label{FigS3}
\end{figure}

Fig.~2 in the main text has shown the precision limit as a function of the perturbation strength for the second-order EP case, where $\delta\theta_{\text{CRB}} \sim \theta^{4}$ for PO threshold operation. Here we further analyze the sensing capability of a third-order EP (EP3) sensor. To realize an EP3, we construct a three-mode system with squeezing in each mode and nearest-neighbor coupling. The Hamiltonian takes the form:
\begin{equation}
    \bm{M} = \begin{pmatrix}
    \bm{M}_1 & \bm{M}_{12} & \bm{M}_{13} \\
    \bm{M}_{12} & \bm{M}_{2} & \bm{M}_{23} \\
    \bm{M}_{13} & \bm{M}_{23} & \bm{M}_{3}
    \end{pmatrix},
\end{equation}
where $\bm{M}_{1,2,3}$ are given by Eq.~(\ref{Mj}), $\bm{M}_{12}=\bm{M}_{23}=\text{diag}\{g, -g, g, -g\}$ and $\bm{M}_{13} = 0$. The system reaches an EP3 and lasing threshold under the condition $g=1/\sqrt{2}$, $\gamma_1=-2$, $\gamma_2=0$, and $\gamma_3=2$ \cite{hodaei2017enhanced}. Moreover, the system reaches an EP3 and PO threshold when $g=1/10\sqrt{2}$, $\epsilon_1=\epsilon_3=2i$, $\epsilon_2=-2i$, $\gamma_1=4.2$, $\gamma_2=4$, and $\gamma_3=3.8$.
Fig.~\ref{FigS3} shows that the precision limit of an EP3 sensor exhibits a $\theta^{3}$ scaling at the lasing threshold, and a $\theta^{6}$ scaling at the PO threshold. These results further verify that the sensitivity enhancement in this scheme arises from the nontrivial interplay between EPs and squeezing.

\subsection{Characteristics of EP sensors at the PO threshold}
We now discuss several additional characteristics of squeezing-enhanced sensors at EPs.

\subsubsection{a.~Photon number dependence}
We evaluate quantum metrology limits by taking into account the resource consumption. Though a $\theta^{2n}$ scaling appears to reduce the precision lower bound for smaller perturbation with no limit, accounting for the number of intracavity photons as the metrology resources reveals that the sensing performance of our sensor still follows the scaling associated with the standard quantum limit. This can be seen from the fact that the leading-order terms in the Green’s function $\bm{G}_{\theta}$ scale as $\theta^{-2n}$ according to Eq.~(\ref{G_scaling}) for a second-order EP sensor at the PO threshold. Since the intracavity field is given by $\bm{a}=\bm{G}_{\theta}\bm{a}_{\rm in}$, the intracavity photon number $N=\langle \bm{a}^{\dagger} \bm{a} \rangle$ scales with the perturbation as $\theta^{-4n}$. It is also noted that QFI scales as $\bm{I}_{\theta}\propto\theta^{-4n}$, proportional to the square of the sensitivity. Therefore, the EP sensor at the PO threshold has $\bm{I}_{\theta}\propto N$ and sensitivity $\propto \sqrt{N}$, consistent with the scaling relation associated with the standard quantum limit. In comparison, the Heisenberg scaling has $\bm{I}_{\theta}\propto N^2$ and sensitivity $\propto N$, which can be further approached if a squeezed-state input is distributed to each input port or the dynamics of state evolution is measured. Here this scheme focuses on the linear scattering properties of squeezing-enhanced EP sensors with coherent or thermal input, without relying on any squeezed probe or dynamic detection protocols. Even so, the sensing scheme exhibits a unique $\theta^{2n}$ scaling in the precision limit, thus offering a superior capability for detecting ultra-weak perturbations.

\subsubsection{b. Noise properties}
\begin{figure}[!htb] 	\centering\includegraphics[width=0.46\textwidth]{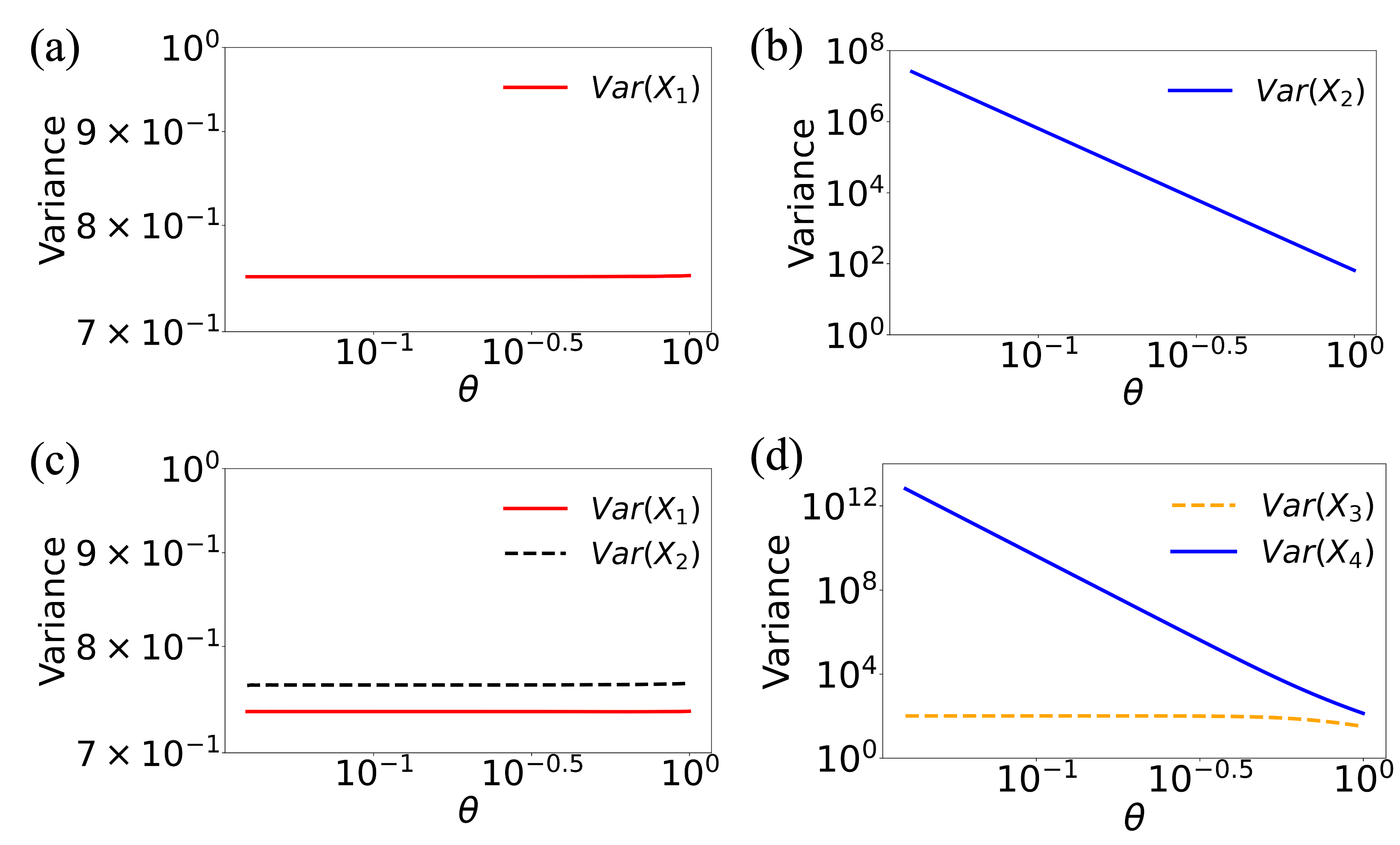}
	\caption{Variances of the quadratures as a function of the perturbation strength for sensors at the PO threshold. (a) Squeezed quadrature ($X_{1}$) and (b) anti-squeezed quadrature ($X_{2}$) of a single-mode sensor. (c) Squeezed quadratures ($X_{1,2}$) and (d) anti-squeezed quadratures ($X_{3,4}$) of a second-order EP sensor. Parameters for the single-mode sensor: $\epsilon=2$, $\gamma_{c} = 1$, $\gamma_{0}=3$. Parameters for the EP sensor: $g=0.1$, $\epsilon_1=2i$, $\epsilon_2=-2i$, $\gamma_{c1} = \gamma_{c2} = 1$, $\gamma_{01}=2.8$, $\gamma_{02}=3.2$.}
\label{FigS4}
\end{figure}

Squeezing is able to reduce the variance of certain quadratures below the shot-noise level, a hallmark quantum feature of the output field. With our quadrature definitions $q=(a+a^\dagger)$ and 
$p=-i(a-a^\dagger)$, the shot-noise level corresponds to a variance of 1. To verify the quantum noise behavior, we numerically evaluate the quadrature variances for both the single-mode and second-order EP sensors at the parametric oscillation (PO) threshold (Fig.~\ref{FigS4}). For each perturbation, we diagonalize the covariance matrix of the output field to obtain the variances of the squeezed and anti-squeezed quadratures. Notably, the covariance matrix eigenvalues appear in degenerate pairs due to the full sideband quadrature representation of the output field ($V_{out}$ is $4\times4$ for the single-mode case and $8\times8$ for the two-mode case). We show only the distinct variances, with each curve representing a doubly degenerate branch. For both cases, the squeezed quadratures exhibit sub-shot-noise behavior [Fig.~\ref{FigS4}(a)(c)]. Moreover, the variances of the squeezed quadratures remain invariant with $\theta$ when $\theta$ is sufficiently small, in contrast to the scaling behavior of anti-squeezed quadrature variances [Fig.~\ref{FigS4}(b)(d)]. These quantum statistical features can be detected using the homodyne and heterodyne detection schemes discussed in Sec.~V.

\subsubsection{c.~Dynamic range}
The dynamic range refers to the range of $\theta$ values that can be precisely estimated. Since the precision limit decreases quartically with $\theta$, the scheme exhibits no lower bound for the detectable dynamic range. The upper limit of the dynamic range is primarily determined by the constraint that $\theta$ is much smaller than the squeezing amplitudes $|\epsilon|$, which is the approximation used for deriving the eigenspectrum near the PO threshold, as shown in Eq.~(\ref{eigenvalues}). This constraint still applies to each unit in a sensor composed of multiple modes, and does not depend on the order of the EP.

\subsubsection{d.~Bandwidth}
The bandwidth of the sensor which sets the timescale for the system to reach steady states, is mainly determined by the poles of the scattering matrix. For an $n$th-order EP sensor operating near the PO threshold, the bandwidth is significantly influenced by the poles near the real axis, while the effect of other poles far away from the real axis on the bandwidth is relatively small and negligible. Given the fact the eigenvalues corresponding to the amplified eigenstates scales as $\theta^2$, the transmission element in the scattering matrix near the zero detuning takes the form 
\begin{equation}
    t(\omega) \approx \frac{A}{\left( \omega - \frac{i\theta^{2}}{2|\epsilon|} \right)^{n}}.
\end{equation}
The half-width at half-maximum (HWHM) is obtained by solving
\begin{equation}
    \left|t(\omega_{\text{HWFM}})\right|^2 \approx \frac{|A|^{2}}{\left( \omega_{\text{HWFM}}^{2} + \frac{\theta^{4}}{4|\epsilon|^{2}} \right)^{n}}=\frac{|A|^{2}}{2\left( \frac{\theta^{4}}{4|\epsilon|^{2}} \right)^{n}}.
\end{equation} 
Therefore, the bandwidth is given by \begin{equation}
    \omega_{\text{HWFM}} = \sqrt{2^{1/n} - 1} \, \frac{\theta^{2}}{2|\epsilon|^{2}}.
\end{equation}
The higher-order EP further amplifies the signal response of the sensor and reduces its bandwidth. It is worth noting that near the PO threshold, the decay of one of the amplified eigenstate approaches zero, leading to critical slowing down of the modes’ response to the variation of the environment. Such critical slowing down phenomenon also exists in other schemes operate near the PO threshold \cite{di2023critical,alushi2024optimality,gorecki2025interplay,beaulieu2025criticality}.

\section{Classical Fisher Information}
We now investigate the precision limit achieved by practical quantum measurement schemes. For a Gaussian output state with mean ${\bm\mu}_{\theta}$ and
covariance ${\bm V}_{\theta}$, the precision lower bound is evaluated by the classical Fisher Information (CFI) given by
\begin{align}
I(\theta) &= \left( \frac{d\, \bm{\mu}_\theta}{d\theta} \right)^T 
\bm{\Sigma}_\theta^{-1} 
\left( \frac{d\, \bm{\mu}_\theta}{d\theta} \right) \nonumber\\
&\quad + \frac{1}{2} \, \mathrm{Tr} \left( 
\bm{\Sigma}_\theta^{-1} \frac{d \bm{\Sigma}_\theta}{d\theta} 
\bm{\Sigma}_\theta^{-1} \frac{d \bm{\Sigma}_\theta}{d\theta} 
\right) \nonumber\\
&= I_\mu(\theta) + I_V(\theta).
\end{align}
For heterodyne measurement, the shot noise adds an identity contribution to each quadrature
\begin{equation}
{\bm\Sigma}_{\!\theta}^{(\mathrm{het})}
    = {\bm V}_{\theta}+{\bm I}_{4}.
\end{equation}
For homodyne measurement, only one rotated quadrature is recorded, with the corresponding covariance matrix expressed as
\begin{equation}
\bm{\Sigma}_{\theta}^{(hom)} = \bm{\Pi} \bm{N}(\phi) \bm{T} \bm{V}_{\theta} \bm{T}^T \bm{N}(\phi)^T \bm{\Pi}.
\end{equation}
Here a transformation matrix
\begin{align}
\bm{\mathrm{T}} = \frac{1}{\sqrt{2}} 
\begin{bmatrix}
1 & 1 & 0 & 0 \\
0 & 0 & 1 & 1
\end{bmatrix},
\end{align}
describes the mapping from the four field quadratures $(q[\omega],q[-\omega],p[\omega],p[-\omega])^{\mathsf T}$ to the two photocurrents $\bigl(q[\omega]+q[-\omega],\;p[\omega]+p[-\omega]\bigr)^{\mathsf T}$. A rotation matrix, 
\begin{align}
\bm{N}(\phi) =
\begin{bmatrix}
\cos\phi & -\sin\phi \\
\sin\phi & \cos\phi
\end{bmatrix},
\end{align}
determines the angle ($\phi$) of measurement in the phase space, which is tunable through the phase of the local oscillator. A projection matrix
\begin{align}
\bm{\Pi} =
\begin{bmatrix}
1 & 0 \\
0 & 0 \\
\end{bmatrix},
\end{align}
indicates that each round of measurement records only one quadrature while discarding the conjugate one. 

The same framework can be extended to the multi-mode bosonic sensing scheme, which performs homodyne or heterodyne detection on the quadratures from each output channel. 

\section{Effect of imperfect PO threshold}
\begin{figure}[!htb] 	\centering\includegraphics[width=0.46\textwidth]{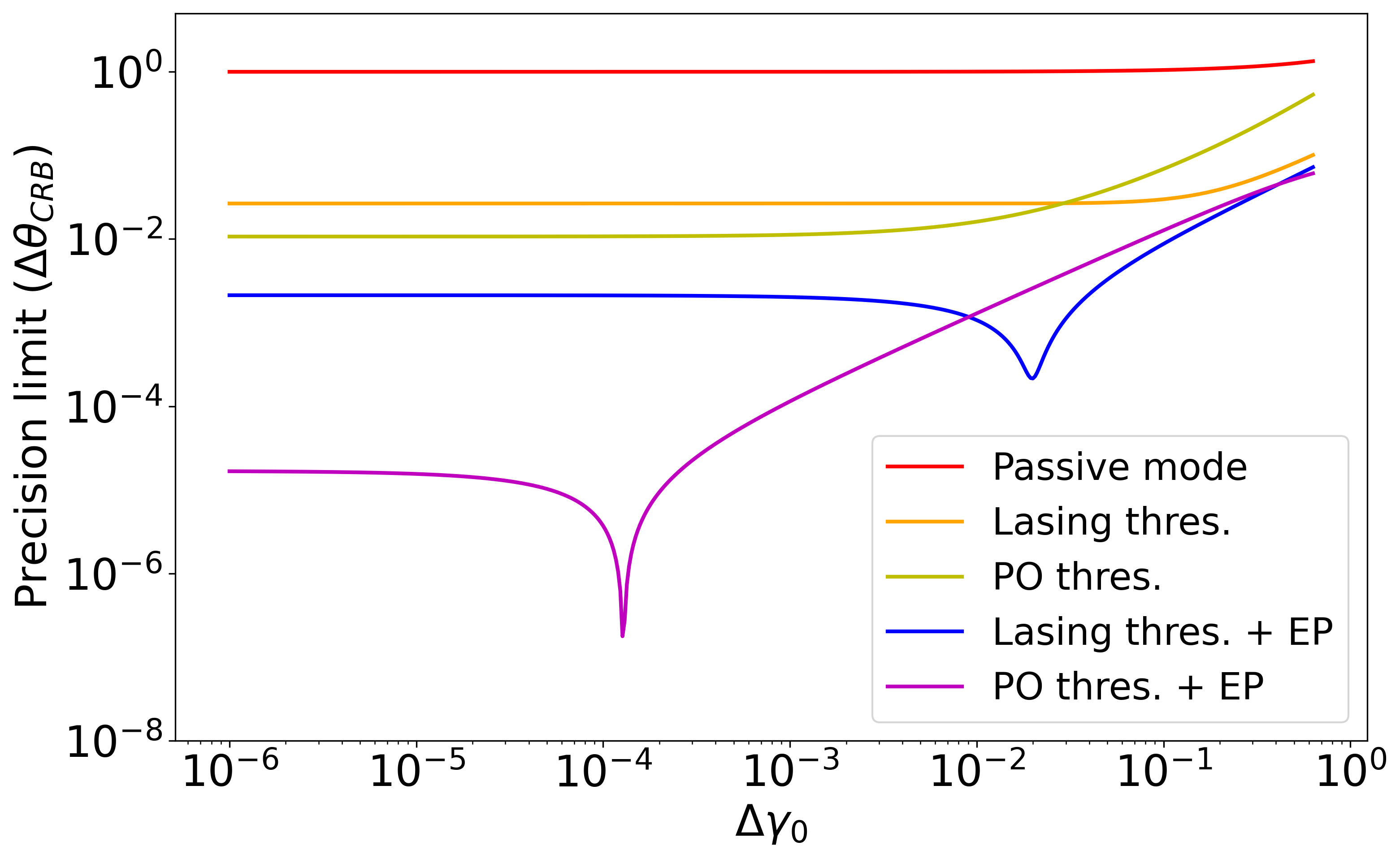}
	\caption{Precision limit as a function of the variation of the intrinsic losses from the lasing/PO threshold $\Delta\gamma_0$ (applied to $\gamma_0$, $\gamma_{01}$, and $\gamma_{02}$), for $\Delta\theta=0.1$.}
\label{FigS1}
\end{figure}

For the sensors operating slightly below the threshold, we predict that the sensitivity enhancement can still be found if the deviation from the threshold is much smaller than the perturbation strength. Fig.~\ref{FigS1} shows that the precision limit derived from QFI as a function of the additional loss added to sensors at the lasing or PO threshold for single-mode and second-order EP sensing schemes. We find that the EP sensor near the PO threshold outperforms other schemes for $\theta=0.1$ when the loss imbalance is below 0.01.

\section{Physical Implementation}
\subsection{Experimental implementation in coupled photonic resonators}
We discuss a practical realization of squeezing-enhanced EP sensing using coupled silicon-nitride (SiN) photonic microring resonators, enabled by the state-of-the-art techniques in contemporary photonic experiments. With current advances in nanophotonic fabrication and quantum metrology, it is feasible to fabricate individual SiN photonic resonator with intrinsic $Q$ factors in the range of $10^6-10^7$ for optical squeezing generation \cite{zhao2020near}. By injecting a two-tone pump into two sideband modes at $\omega_0\pm\delta\omega$, single-mode squeezing at the center resonance $\omega_0$ can be generated concurrently through the degenerate four wave mixing process. To engineer a second-order EP, one can fabricate resonators on edges of separate chips by undercutting the substrate and coupling the two resonators with tunable coupling strength via spacing their gap using piezoelectric stages \cite{peng2014parity, wang2021coherent}. The resonance frequencies of the two resonators can be aligned via electro-thermal tuning. The pump is split and injected to each waveguide coupled to the resonator, with the phases and amplitudes properly adjusted \cite{wang2021coherent}. 

\begin{figure}[!htb] 	\centering\includegraphics[width=0.46\textwidth]{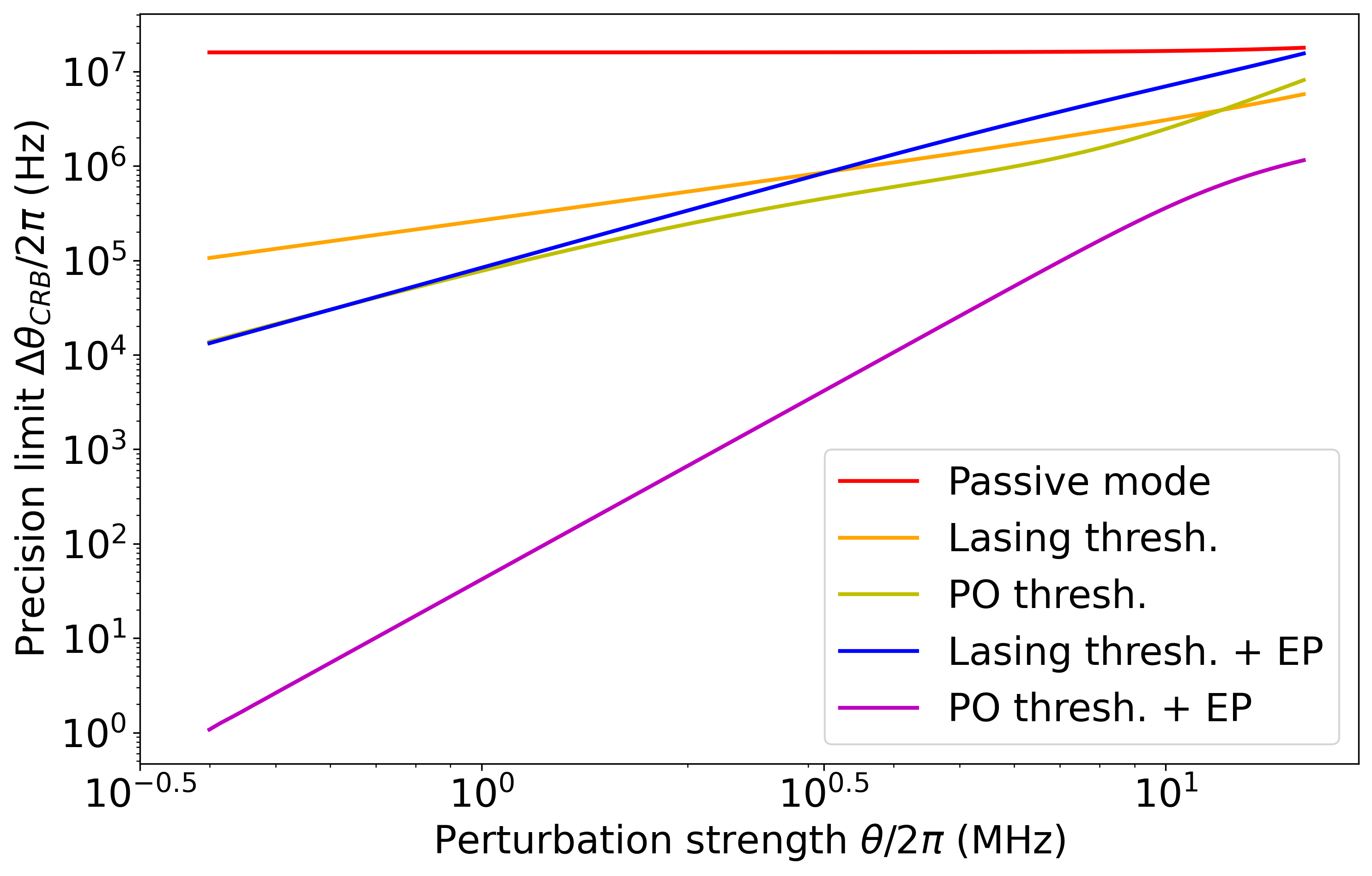}
	\caption{Precision limit (Cramér-Rao bound) as a function of the perturbation strength for different cases under realistic parameters. For the coupled microring resonators supporting a parametric-oscillation threshold at a second-order EP, the following parameters are used based on state-of-the-art SiN photonic devices: intrinsic losses $\gamma_{01}=320$ MHz, $\gamma_{02}=280$ MHz; external losses $\gamma_{c1,c2}=100$ MHz; effective squeezing amplitudes $\epsilon_{1}=200i$ MHz, $\epsilon_{2}=-200i$ MHz; and inter-resonator coupling strength $g=10$ MHz.}
\label{FigS2}
\end{figure}

Fig.~\ref{FigS2} shows the numerical evaluation of the Cramér-Rao bound for all the cases listed in the main text by applying realistic parameters of state-of-the-art SiN photonic devices.

Beyond these quantitative estimates, we discuss the key technical challenges and strategies for experimental implementation using SiN photonic devices. The main difficulty lies in maintaining stable operation while varying the perturbation strength, as parameter fluctuations can drive the system away from the optimal operating point. For instance, mechanical vibrations may alter the inter-resonator coupling, while temperature drifts can detune the resonances and steer the system away from the EP. As system complexity grows, manual tuning becomes impractical. A viable solution is to implement an automated feedback control system \cite{dong2019sensitive} capable of performing real-time fitting of the transmission spectrum to extract and monitor all the system parameters, while providing real-time feedback to the piezo, electro-thermal, and pump controls. Such an automated control system locking the system to an EP and PO threshold would substantially improve stability and potentially make it feasible to operate multimode sensors at higher-order EPs. Additionally, exceptional surfaces, a hypersurface of EPs in a high-dimensional parameter space, provide an alternative strategy to improve the robustness of EP systems against parameter fluctuation \cite{zhong2019sensing}.  

\subsection{Unidirectional coupling scheme}
Beyond the foregoing scheme, the squeezing-enhanced EP sensing can also be implemented using a unidirectional coupling scheme. Let us consider two modes with the same frequencies both operating at the PO threshold. Assume there is coupling only from the first mode to the second mode with no reflection. Such a scheme has been realized in whispering gallery mode optical resonators where the coupling between clockwise and counterclockwise modes is manipulated by a nano scatterer \cite{chen2017exceptional, wang2020electromagnetically, wang2021induced}. The Hamiltonian of the system under the quadrature basis is given by 
\begin{equation}
    M =
\begin{pmatrix}
M_1 & M_{12} \\
M_{21} & M_2
\end{pmatrix},
\end{equation}
where $M_j$ is given by Eq.~(\ref{Mj}), $M_{21}=\text{diag}\{g, -g, g, -g\}$, and $M_{12} = 0$. The determinant takes the form 
\begin{equation}
\Delta = \frac{1}{256}
\left( 4\theta^{2} + \gamma_{1}^{2} - 4|\epsilon_{1}|^{2} \right)^{2}
\left( 4\theta^{2} + \gamma_{2}^{2} - 4|\epsilon_{2}|^{2} \right)^{2}.
\end{equation} 
At the PO threshold, $\Delta = \frac{1}{64} \, \theta^{8}$. By calculating the eigenspectrum and eigenstates, we find that there exist two pairs of degenerate eigenstates with corresponding eigenvalues scaling as $\theta^2$, indicating the presence of a second-order EP. Furthermore, such a cascaded scheme can be extended to an nth-order EP at the PO threshold, leading to the $\theta^{2n}$ precision scaling. This example broadens the routes for engineering the EPs in systems with squeezing to achieve enhanced sensing.

\bibliographystyle{apsrev4-2}
\bibliography{SI_references}